\documentclass[sigconf,screen,natbib=false]{acmart}
\pdfoutput=1
\renewcommand\footnotetextcopyrightpermission[1]{} 

\usepackage[utf8]{inputenc}
\usepackage{latexsym}
\usepackage{amsmath}
\usepackage{amsthm}
\usepackage{thm-restate}
\usepackage{mathtools}
\usepackage{newtxtext,newtxmath}
\usepackage{graphicx}
\usepackage{ifthen}
\usepackage{calc}
\usepackage{MnSymbol}
\usepackage{enumitem}

\usepackage{BOONDOX-cal}

\usepackage[capitalize]{cleveref}
\usepackage[thinc]{esdiff}

\usepackage[
  datamodel=acmdatamodel,
  style=acmnumeric, 
  ]{biblatex}

\DeclareSourcemap{
 \maps[datatype=bibtex]{
   \map{
     \step[fieldsource=url,
       match=\regexp{\\_},
       replace=\regexp{_}]
   }
 }
}

\renewbibmacro*{doi+eprint+url}{\iftoggle{bbx:url}
  {\iffieldundef{doi}{\usebibmacro{url+urldate}}{}}
  {}\newunit\newblock
  \iftoggle{bbx:eprint}
  {\usebibmacro{eprint}}
  {}\newunit\newblock
  \iftoggle{bbx:doi}
  {\printfield{doi}}
  {}}

\usepackage{tikz}

\allowdisplaybreaks

\makeatletter

\newsavebox{\tempbox}
\renewcommand{\@makecaption}[2]{
  \vspace{10pt}
  \sbox{\tempbox}{\textbf{#1.} #2}
  \ifthenelse{\lengthtest{\wd\tempbox > \linewidth}}{
    \textbf{#1.} #2\par
  }{
    \begin{center}
      \textbf{#1.} #2
    \end{center}
  }
}

\makeatother

\numberwithin{equation}{section}
\renewcommand{\theequation}{\arabic{section}.\Alph{equation}}
  
\numberwithin{figure}{section}

\newtheoremstyle{mythm}
  {}
  {}
  {\itshape}
  {}
  {\bfseries}
  {.}
  {.5em}
  {\thmname{#1}~\thmnumber{#2}\ifthenelse{\equal{\thmnote{#3}}{}}{}{~(\thmnote{#3})}}

\newtheoremstyle{mydefn}
  {}
  {}
  {\upshape}
  {}
  {\bfseries}
  {.}
  {.5em}
  {\thmname{#1}~\thmnumber{#2}\ifthenelse{\equal{\thmnote{#3}}{}}{}{~(\thmnote{#3})}}

\newtheoremstyle{myremark}
  {}
  {}
  {\upshape}
  {}
  {\itshape}
  {.}
  {.5em}
  {\thmname{#1}~\thmnumber{#2}\ifthenelse{\equal{\thmnote{#3}}{}}{}{~(\thmnote{#3})}}

\theoremstyle{mythm}
\newtheorem{theorem}{Theorem}[section]
\newtheorem{lemma}[theorem]{Lemma}

\newtheorem{corollary}[theorem]{Corollary}

\theoremstyle{mydefn}

\newtheorem{example}[theorem]{Example}
\theoremstyle{myremark}
\newtheorem{remark}[theorem]{Remark}
\theoremstyle{mythm}

\newcommand{\uend}{\hfill$\lrcorner$}

\newcounter{claimcounter}
\newenvironment{claim}[1][]{
  \renewcommand{\proof}{\smallskip\par\noindent\textit{Proof. }}
  \medskip\par\noindent%
  \ifthenelse{\equal{#1}{}}{%
    \setcounter{claimcounter}{0}\refstepcounter{claimcounter}\textit{Claim~\arabic{claimcounter}.}
  }{%
    \ifthenelse{\equal{#1}{resume}}{%
      \refstepcounter{claimcounter}\textit{Claim~\arabic{claimcounter}.}
    }{%
      \textit{Claim~#1.}
    }
  }
}{
  \par\medskip
}

\newlist{caselist}{description}{10}
\setlist[caselist]{font=\itshape\mdseries}

\setenumerate[1]{label=(\arabic*)}
\newlist{eroman}{enumerate}{2}
\setlist[eroman,1]{label=(\roman*)}
\setlist[eroman,2]{label=(\alph*)}
\newlist{ealph}{enumerate}{1}
\setlist[ealph]{label=(\Alph*)}

\newcounter{nlistcounter}

\usepackage{color}
\definecolor{blau}{RGB}{0,84,159}
\definecolor{hellblau}{RGB}{142,168,229}
\definecolor{petrol}{RGB}{0,97,101}
\definecolor{tuerkis}{RGB}{0,152,161}
\definecolor{gruen}{RGB}{87,171,39}
\definecolor{maigruen}{RGB}{189,205,0}
\definecolor{gelb}{RGB}{255,237,0}
\definecolor{orange}{RGB}{255,128,0}
\definecolor{magenta}{RGB}{227,0,102}
\definecolor{rot}{RGB}{204,7,30}
\definecolor{bordeaux}{RGB}{161,16,53}
\definecolor{violett}{RGB}{97,33,88}
\definecolor{lila}{RGB}{122,111,172}
\definecolor{grey}{gray}{0.7}
\definecolor{mittelblau}{RGB}{0,128,255}

\newcommand{\alert}[1]{\ifmmode{#1}\else{\emph{#1}}\fi}

\usepackage[textwidth=1.7cm,textsize=footnotesize]{todonotes}
\makeatletter
\@mparswitchfalse%
\normalmarginpar
\tikzstyle{notestyleraw} = [
    draw = none,
    fill = \@todonotes@currentbackgroundcolor,
    text = \@todonotes@currenttextcolor,
    line width = 1pt,
    text width = \@todonotes@textwidth - 1.6 ex - 2pt, 
    inner sep = 0.8 ex,
    xshift = 0cm
    ]
\makeatother

\makeatletter
\def\@er{ER}
\def\@mg{MG}
\makeatother

\makeatletter
\newcommand*{\s@}{}
\newcommand*{\s@er}{\s@\setuptodonotes{color=orange,textcolor=black,caption={}}}
\newcommand*{\s@mg}{\s@\setuptodonotes{color=red,textcolor=black,caption={}}}
\makeatother

\makeatletter
\newcommand{\ier}[2][\@nil]{\s@er\todo[inline]{\sffamily\textbf{\@er\def\tmp{#1}\ifx\tmp\@nnil\else~[#1]\fi:~}#2}}
\newcommand{\eran}[2][\@nil]{\s@bk\todo        {\sffamily\textbf{\@bk\def\tmp{#1}\ifx\tmp\@nnil\else~[#1]\fi:~}#2}}
\newcommand{\imartin} [2][\@nil]{\s@mg\todo[inline]{\sffamily\textbf{\@mg\def\tmp{#1}\ifx\tmp\@nnil\else~[#1]\fi:~}#2}}
\newcommand{\martin}[2][\@nil]{\s@mg\todo        {\sffamily\textbf{\@mg\def\tmp{#1}\ifx\tmp\@nnil\else~[#1]\fi:~}#2}}
\makeatother

\newcommand{\bigmid}{\mathrel{\big|}}
\newcommand{\Bigmid}{\mathrel{\Big|}}
\newcommand{\ceil}[1]{\left\lceil#1\right\rceil}
\newcommand{\floor}[1]{\left\lfloor#1\right\rfloor}
\newcommand{\angles}[1]{\left\langle#1\right\rangle}
\renewcommand{\tilde}{\widetilde}
\renewcommand{\hat}{\widehat}

\renewcommand{\vec}[1]{\boldsymbol{#1}}

\newcommand{\lmulti}{{\{\hspace{-2.0pt}\{}}
\newcommand{\rmulti}{{\}\hspace{-2.0pt}\}}}
\newcommand{\biglmulti}{{\big\{\hspace{-2.8pt}\big\{}}
\newcommand{\bigrmulti}{{\big\}\hspace{-2.8pt}\big\}}}
\newcommand{\Biglmulti}{{\Big\{\hspace{-3.6pt}\Big\{}}
\newcommand{\Bigrmulti}{{\Big\}\hspace{-3.6pt}\Big\}}}
\def\multi#1{\lmulti#1\rmulti}
\def\multiset#1#2{\ensuremath{\left(\kern-.3em\left(\genfrac{}{}{0pt}{}{#1}{#2}\right)\kern-.3em\right)}}
\def\abs#1{\big|#1\big|}

\renewcommand{\phi}{\varphi}
\renewcommand{\epsilon}{\varepsilon}

\newcommand{\Nat}{{\mathbb N}}
\newcommand{\PNat}{{\mathbb N}_{>0}}
\newcommand{\Real}{{\mathbb R}}

\newcommand{\LL}{\textsf{\upshape L}}
\newcommand{\LC}{\textsf{\upshape C}}
\newcommand{\FO}{\textsf{\upshape FO}}

\newcommand{\CF}{{\mathcal F}}
\newcommand{\CG}{{\mathcal G}}

\newcommand{\CM}{{\mathcal M}}
\newcommand{\CN}{{\mathcal N}}

\newcommand{\CQ}{{\mathcal Q}}

\newcommand{\CS}{{\mathcal S}}

\newcommand{\CX}{{\mathcal X}}

\newcommand{\Cb}{\mathcal b}
\newcommand{\Cf}{\mathcal f}
\newcommand{\Cg}{\mathcal g}

\newcommand{\CGS}{{\mathcal G\!\mathcal S}}

\newcommand{\FN}{{\mathfrak N}}
\newcommand{\FL}{{\mathfrak L}}

\newcommand{\Fm}{{\mathfrak w}}

\newcommand{\bool}{{\textup{bool}}}
\DeclareMathOperator{\sig}{sig}

\DeclareMathOperator{\relu}{relu}
\DeclareMathOperator{\id}{id}

\DeclareMathOperator{\Bit}{Bit}
\DeclareMathOperator{\sign}{sign}

\newcommand{\FC}{\mathcal{FC}}
\newcommand{\FCco}{\mathcal{FC}^{\textup{co}}}

\newcommand{\logic}[1]{\textsf{\upshape #1}}

\newcommand{\MC}[1][]{\logic{MC}}
\newcommand{\GC}[1][]{\logic{GC}}
\newcommand{\FOC}[1][]{\logic{FO$^{#1}$+C}}
\newcommand{\MFOC}{\logic{MFO+C}}
\newcommand{\GFOC}{\logic{GFO+C}}
\newcommand{\TC}{\logic{TC}}

\newcommand{\ord}{\logic{ord}}
\newcommand{\bit}{\logic{bit}}

\newcommand{\Fp}{{\mathfrak p}}

\newcommand{\num}[1]{\left\llangle#1\right\rrangle}

\newcommand{\set}[1]{\left[#1\right[}

\newcommand{\agg}{\logic{agg}}
\newcommand{\comb}{\logic{comb}}
\newcommand{\msg}{\logic{msg}}
\newcommand{\Msg}{\logic{Msg}}

\newcommand{\SUM}{\logic{SUM}}
\newcommand{\MEAN}{\logic{MEAN}}
\newcommand{\MAX}{\logic{MAX}}
\newcommand{\mean}{\logic{mean}}

\newcommand{\nuni}{_{\textup{nu}}}

\begin{document}
\title{Are Targeted Messages More Effective?}
\author{Martin Grohe}
\authornote{Funded by the European Union (ERC, SymSim,
101054974). Views and opinions expressed are however those of the author(s) only and do not necessarily reflect those of the European Union or the European Research Council. Neither the European Union nor the granting authority can be held responsible for them.}
\orcid{orcid.org/0000-0002-0292-9142}
\author{Eran Rosenbluth}
\authornote{Funded by the Deutsche Forschungsgemeinschaft (DFG) - 2236/2.}
\orcid{orcid.org/0000-0002-5629-2220}
\email{{grohe, rosenbluth}@informatik.rwth-aachen.de}
\affiliation{
   \institution{RWTH Aachen University} \country{Germany}
}

\begin{abstract}  
  Graph neural networks (GNN) are deep learning architectures for
  graphs. Essentially, a GNN is a distributed message passing
  algorithm, which is controlled by parameters learned from data. It operates
  on the vertices of a graph: in each iteration, vertices receive a message on each incoming edge, aggregate these messages, and then update
  their state based on their current state and the aggregated messages.
  The expressivity of GNNs can be characterised in terms of certain
  fragments of first-order logic with counting and the
  Weisfeiler-Lehman algorithm.

   The core GNN architecture comes in two different
   versions. In the first version, a message only depends on the state of the source vertex, whereas in the second version it depends on the states of the source and target vertices.
   In practice, both of these
   versions are used, but the theory of GNNs so far mostly focused
   on the first one. On the logical side, the two versions correspond to
   two fragments of first-order logic with counting that we call modal and guarded.
   
   The question whether the two versions differ in their expressivity
   has been mostly overlooked in the GNN literature and has only been
   asked recently (Grohe, LICS'23). We answer this question here. It
   turns out that the answer is not as straightforward as one might
   expect. By proving that the modal and guarded fragment of
   first-order logic with counting have the same expressivity over
   labelled undirected graphs, we show that in a non-uniform setting
   the two GNN versions have the same expressivity. However, we also
   prove that in a uniform setting the second version  is strictly more expressive. 
 \end{abstract}


\maketitle 
\pagestyle{plain}

\section{Introduction}

The question we address in this paper is motivated by comparing
different versions of graph neural networks,  but turns out to be
related to a very natural question in logic concerning the difference
between modal and guarded fragments of first-order logic with
counting.

\subsection*{Graph Neural Networks}
Among the various deep learning architectures for graphs, message passing graph neural networks (GNNs) \cite{GallicchioM10,GilmerSRVD17,KipfW17,ScarselliGTHM09} are the most common choice for many different applications
ranging from combinatorial optimisation to particle physics
(see, for example, \cite{BernreutherFKKM21,CappartCK00V23,ChamiAPRM22})

We think of a GNN as a distributed message passing algorithm operating
on the vertices of the input graph $G$. This perspective goes back to
Gilmer et al.~\cite{GilmerSRVD17}.  Each vertex $v$ of $G$ has a state
$\Cf(v)$, which is a vector over the reals.\footnote{We might also call the state of a vertex \emph{feature
    vector}; this explains the use of the letter $\Cf$.} In this paper, the input
graphs $G$ are undirected vertex-labelled graphs, and the initial
state $\Cf(v)$ of a vertex $v$ is an encoding of its
labels.  Then in each
iteration all vertices send a message to all their neighbours. Upon
receiving 
the messages from its neighbours, a vertex $v$ aggregates
all the incoming messages, typically using coordinatewise  summation,
arithmetic mean, or maximum, and then updates its state by applying a
\emph{combination function} to its current state and the aggregated
messages. The states of the vertices after the last iteration are used
to compute the output of the GNN. The combination function as well as
the \emph{message function} used to compute the messages are computed
by neural networks whose parameters are learned from
data. Importantly, all vertices use the same message, aggregation, and
combination functions. Thus they share the learned parameters, which
means that the number of parameters is independent of the size of
the input graph. It also means that a GNN can be applied to graphs of
arbitrary size.  Our GNNs carry out a fixed number of iterations, and
each iteration (or \emph{layer} of the GNN) has its own combination
and message functions. This is the most basic model and also the
``standard'', but there is a model of recurrent GNNs (see, for
example, \cite{Grohe23}), which we do
not study here.

Let us be clear that the view of a GNN as a distributed message
passing algorithm is an abstract conceptualisation that we use to
explain how GNNs work. In an actual implementation of GNNs, nobody
sends any messages. A central control instance collects the states of
all vertices in a matrix, and the computation is reduced to matrix and
tensor computations, typically carried out on a GPU using highly
optimised code \cite{wang2019deep, fey2019fast}. (The computation is parallelised, but the
parallelisation is not tied to the initial distributed algorithm.)

\begin{figure}
  \centering
  \begin{tikzpicture}
    \footnotesize
    \begin{scope}
      \node[circle,draw] (u) at (0,0) {$u$};
      \node[circle,draw] (v1) at (90:15mm) {$v_1$};
      \node[circle,draw] (v2) at (330:15mm) {$v_2$};
      \node[circle,draw] (v3) at (210:15mm) {$v_3$};

      \draw[thick,->] (u) edge[bend left] node[left]
      {$\msg(\Cf(u))$} (v1)
      (v1) edge[bend left] node[right] {$\msg(\Cf(v_1))$} (u)
      (u) edge[bend left] (v2) 
      (v2) edge[bend left] (u) 
      (u) edge[bend left] (v3) 
      (v3) edge[bend left] (u) 
      ;
      \path (0,-1.5) node {(a)};
    \end{scope}
    \begin{scope}[xshift=4.5cm]
      \node[circle,draw] (u) at (0,0) {$u$};
      \node[circle,draw] (v1) at (90:15mm) {$v_1$};
      \node[circle,draw] (v2) at (330:15mm) {$v_2$};
      \node[circle,draw] (v3) at (210:15mm) {$v_3$};

      \draw[thick,->] (u) edge[bend left] node[left]
      {$\msg(\Cf(v_1),\Cf(u))$} (v1)
      (v1) edge[bend left] node[right] {$\msg(\Cf(u),\Cf(v_1))$} (u)
      (u) edge[bend left] (v2) 
      (v2) edge[bend left] (u) 
      (u) edge[bend left] (v3) 
      (v3) edge[bend left] (u) 
      ;
      \path (0,-1.5) node {(b)};
    \end{scope}
  \end{tikzpicture}
  \caption{1-sided and 2-sided message passing in the graph
    $G=\big(\{u,v_1,v_2,v_3\},\{uv_1,uv_2,uv_3\}\big)$. Every vertex $x$ has a state
    $\Cf(x)$. Messages are associated with oriented edges. In the
    1-sided version (a), messages are a function of the state of the source vertex, in
    the 2-sided version (b) they are function of the states of both
    source and target.}
\label{fig:1}
\end{figure}
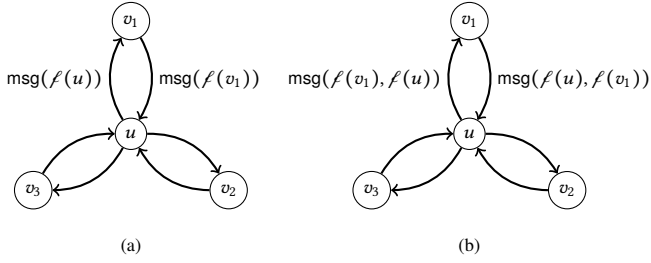

By now, there are numerous variants of the basic message passing model
just described. But even the basic model comes in two different
versions. In a \emph{1-sided GNN} (for short: \emph{1-GNN}), the message sent along
the edge $(v,u)$ is a function of the state $\Cf(v)$ of the source
vertex. In a \emph{2-sided GNN (2-GNN)}, the message is a function of the
states $\Cf(u)$ and $\Cf(v)$ of both source and target vertex (see Figure~\ref{fig:1}). Thus,
potentially 2-GNNs are more powerful because they allow for ``targeted
messages'' depending on properties of the target vertex. In this paper, we
study the question whether 2-GNNs are indeed more powerful.

Both the 1-sided and the 2-sided versions have been studied before, and both have been used in
practice. Instances of 1-GNNs are the graph convolutional networks of
\cite{KipfW17}. The message passing neural networks of \cite{GilmerSRVD17}
are 2-GNNs. Most of the theoretical work on GNNs considers 1-GNNs
(e.g.~\cite{BarceloKM0RS20,MorrisRFHLRG19,XuHLJ19}); this also has to
do with the fact that this work is concerned with non-uniform
expressivity, for which it is easy to see that the two versions
coincide. In practical work, the prevalent perception seems to be that
2-GNNs are superior---this is also our own
experience---though an explicit empirical comparison between the two
versions has only been carried out recently \cite{TailorOLL22} (there
1-GNNs are called isotropic and 2-GNNs are called
anisotropic).
Interestingly, the conclusion of the this empirical
study is that a novel version of 1-GNNs is en-par with 2-GNNs.

We ask if 2-GNNs are more \emph{expressive} than 1-GNNs, that is,
whether they can compute more functions. We are not concerned with
learning, specifically the algorithmic optimisation and statistical
generalisation questions involved. Obviously, these are highly
relevant questions as well, but technically quite different. Expressivity
results such as the characterisation of the power of GNNs to
distinguish two graphs in terms of the Weisfeiler-Leman algorithm
\cite{MorrisRFHLRG19,XuHLJ19} have been quite influential in the
development of GNNs as they steered research in the direction of more
expressive architectures and helped to understand their limitations
(see, for example, \cite{MorrisLMRKGFB23,MorrisDMCFLLBGJ24a}).

Our focus here is on the basic task of node
classification, or \emph{unary queries} in the language of finite
model theory and database theory. We can build on powerful
characterisations of the queries computable by GNNs in terms of
fragments of first-order logic with counting
\cite{BarceloKM0RS20,Grohe23}. The precise question we ask is:
\begin{quote}\itshape
  Are there unary queries computable by 2-GNNs that are not computable
  by 1-GNNs?
\end{quote}
(The question was left as an open problem in \cite{Grohe23}.)  This
straightforward question has a surprisingly complicated answer, which
has to do with the fact that there are different forms of GNN expressivity. The most natural is \emph{uniform
  expressivity}. A query $\CQ$ is \emph{uniformly expressible} by
$i$-GNNs if there is a single $i$-GNN $\FN$ computing the
query. However, much of the literature on the theory of GNNs considers
\emph{non-uniform expressivity}. A query $\CQ$ is \emph{non-uniformly expressible} by $i$-GNNs if there is a family $\CN=(\FN_n)_{n\ge 1}$
of $i$-GNNs such that $\FN_n$ computes $\CQ$ on input graphs of size
$n$.  There is an argument to be made for non-uniform expressivity,
because in many practical applications the input graphs can be
expected to be of similar size, and this size is also well-represented
in the training data. However, in
this case we still want to control the complexity of the GNNs
involved; certainly GNNs of size exponential in the input graphs would
be infeasible. From this perspective, it is natural to consider
families of GNNs of \emph{polynomial size} and \emph{bounded depth}.

\begin{theorem}\label{thm:main-gnn}
  \begin{enumerate}
  \item There is a query uniformly expressible by a 2-GNN with SUM aggregation,
    but not by a 1-GNN with SUM aggregation\/ \textup{(\cref{theo:unisep})}.
  \item All queries non-uniformly expressible by families of 2-GNNs
    are also non-uniformly expressible by families of 1-GNNs\/ \textup{(\cref{cor:nonuniform})}.
  \item All queries non-uniformly expressible by families of 2-GNNs of
    bounded depth and polynomial size
    are also non-uniformly expressible by families of 1-GNNs of
    bounded depth and polynomial size\/  \textup{(\cref{cor:bounded-nonuniform})}.
  \end{enumerate}
\end{theorem}

Assertion (2) is an easy consequence of the characterisation of the
distinguishing power of GNNs in
terms of the Weisfeiler-Leman algorithm
\cite{MorrisRFHLRG19,XuHLJ19}. This has probably been observed by
others. Assertions (1) and (3) are new and
nontrivial. For (1) we give a direct proof. It is based on an invariant
that the states of the vertices preserve during a 1-GNN computation. This
invariant asserts
that the states can be approximated in terms of certain ``nice''
polynomials. To prove (3), we rely on a logical characterisation of
GNN-expressivity and then prove that the logics characterising
1-GNNs and 2-GNNs coincide. We do not have a direct proof of (3) that
does not go through logic.

Interestingly, it is necessary to use SUM-aggregation in assertion (1) of the theorem: we prove that for GNNs that only use \MAX-aggregation or only use \MEAN-aggregation, 1-GNNs and 2-GNNs have the same expressivity (\cref{thm:mean-max}).

\subsection*{Logic}
1-sided and 2-sided message passing in GNNs correspond very naturally
to two forms of local quantification in logic that appear in modal
and guarded logics.

We continue to consider undirected labelled
graphs, and we interpret all logics over such graphs. The two
forms of quantification can best be explained using the extension $\LC$ of first-order logic by counting quantifiers
$\exists^{\ge n}x.\psi$ (``there exist at least $n$ vertices
$x$ such that $\psi$ holds''). Consider the following two forms of restricted quantification
over the neighbours of a vertex $x$, both saying that there exist at
least $n$ neighbours $x'$ of 
$x$ such that $\psi$ holds:
\begin{gather}
  \label{eq:modal}
  \exists^{\ge n} x'\big(E(x,x')\wedge \psi(x')\big),\\
  \label{eq:guarded}
  \exists^{\ge n} x'\big(E(x,x')\wedge \psi(x,x')\big).
\end{gather}
Importantly, $x$ is not a free variable of
$\psi$ in \eqref{eq:modal}, but it is in \eqref{eq:guarded}. This is
the only difference between \eqref{eq:modal} and \eqref{eq:guarded}. We
call the first form of quantification \emph{modal} and the second
\emph{guarded}. We define the \emph{modal fragment} $\MC$ of $\LC$ to
consist of all $\LC$-formulas that only use two variables and only
modal quantification of the form \eqref{eq:modal}. Similarly, we
define the \emph{guarded fragment} $\GC$ using guarded quantification
of the form \eqref{eq:guarded}.

Intuitively, both the modal and the guarded fragments are
\emph{local}, in the sense that the definable properties of a vertex $u$
depend on the properties of its neighbours $v$. Vertices have no
identities, so the neighbours can only be distinguished in terms of
the properties they have. This means that the properties of $u$ can
only depend on how many neighbours of each property it has. From a
more operational perspective, we can think of modal and guarded quantification as
follows: each neighbour $v$ sends a messages $M_\psi$ to $u$
indicating if it has some property $\psi(v)$. Node $u$ aggregates
these messages by counting how many neighbours satisfy $\psi$ and makes
a decision based on this count. The difference between modal
and guarded quantification is that in guarded quantification the
message $M_\psi$ not only depends on properties of the source $v$, but
on combined properties $\psi(u,v)$ of the source $v$ and the target $u$. That is, $v$ does not indiscriminately send the the same message
to all of its neighbours, but instead sends a tailored message that
depends on the properties of the target $v$ as well. In this sense,
modal and guarded quantification resemble message passing in 1-sided
and 2-sided GNNs.

It is fairly easy to see that the modal fragment $\MC$ and the guarded
fragment $\GC$ have the same expressive power (recall that we are
interpreting the logics over undirected graphs). However, this changes
if we consider the modal fragments $\MFOC$ and the guarded fragment
$\GFOC$ of the stronger logic $\FOC$. In the logic $\LC$, the number
$n$ in a  quantifier $\exists^{\ge n}$ is a constant that is
hard-wired into a formula. In $\MFOC$, the constant $n$ is replaced by
a variable
that ranges over the natural numbers. This allows us, to compare
counts of arbitrary size and to express arithmetic properties of
counts. For example, we can express that a vertex has even degree, which is impossible in
$\LC$. Formally, instead of quantifiers $\exists^{\ge n}x.\psi$ we use
terms $\#x.\psi$, interpreted as the number of $x$ such that $\psi$ is
satisfied. Then the following formula expresses that $x$ has even
degree.
\begin{equation}
  \label{eq:7}
  \exists y<\ord.\Big(2\cdot y =\# x'.E(x,x')\Big).
\end{equation}
Here $y$ is a \emph{number variable} ranging over the natural numbers
and $x,x'$ are \emph{vertex variables} ranging over the vertices of a
graph. We allow quantification over the natural numbers, but only
bounded quantification. In the formula above, we use the constant
$\ord$, which is always interpreted by the order of the input graph,
to bound the existential quantification over $y$.
In a nutshell, $\FOC$ combines first-order logic over finite graphs
with bounded arithmetic, connecting them via counting terms. The syntax and semantics of $\FOC$ and its fragments will be discussed
in detail in Section~\ref{sec:counting-logics}.

In the modal fragment $\MFOC$ and the guarded fragment $\GFOC$ we restrict counting terms in the
way as we restrict quantification in \eqref{eq:modal} and
\eqref{eq:guarded}, respectively.

The logics $\MFOC$ and
$\GFOC$ are interesting because they precisely characterise the
non-uniform expressivity of families of $1$-GNNs and $2$-GNNs, respectively, of
polynomial size and bounded depth \cite{Grohe23}.

It is not obvious at all whether $\MFOC$ and $\GFOC$ have the same
expressive power.
The following example illustrates a sequence of
increasingly harder 
technical issues related to this question, and how we resolve them.

\begin{example}\label{exa:intro}
  \begin{enumerate}
    \item Consider the query $\CQ_1$ selecting all vertices of a graph
    that have a neighbour of larger degree. The query can easily be
    expressed by the $\GFOC$-formula
    \begin{equation}
      \label{eq:6}
       \phi_1(x)\coloneqq \exists
      x'\big(E(x,x')\wedge\logic{deg}(x)<\logic{deg}(x')\big),
    \end{equation}
    where $\logic{deg}(x)$ is the term $\# x'.\big(E(x,x')\wedge
    x'=x'\big)$, which is even modal. Incidentally, it is also easy to
    design a 2-GNN that computes the query $\CQ_1$.

    In Section~\ref{sec:uniform}, we shall prove that the query is not
    computable by a 1-GNN; not even on complete bipartite graphs
    $K_{m,n}$.

    However, the query is expressible in
    $\MFOC$ by the following formula:
    \begin{equation}
      \label{eq:8}
      \exists y<\ord.\Big(y=\logic{deg}(x)\wedge\exists
    x'.\big(E(x,x')\wedge y<\logic{deg}(x')\big)\Big).
    \end{equation}
    We exploit the fact that in the logics $\MFOC$
    and $\GFOC$, only quantification over vertex variables $x'$ is
    restricted to the neighbours of the current vertex $x$, whereas
    quantification over number variables $y$ is not subject to any
    guardedness restrictions. The
    philosophy behind this is that we want the logics to be local in
    the input graph, but want to include full bounded arithmetic.
    
    \item The trick in (1) is to pass information
    across modal quantifiers using number variables. But since every
    formula only contains finitely many number variables, we can
    only pass finitely many numbers. Therefore, the natural idea to
    separate the logics would be to consider properties of nodes that
    depend on unbounded sets of numbers.

    Let $\CQ_2$ be the query that selects all nodes $x$ that have a
    neighbour $x'$ such that sets of degrees of the neighbours of $x$
    and $x'$, respectively, are equal. That is, $\CQ_2$ evaluates to
    true at a node $v$ of a graph $G$ if and only if $v$ has a
    neighbour $v'$ such that
    \begin{equation}
      \label{eq:5}
      \big\{\deg_G(w)\bigmid w\in N_G(v) \big\}
      =
      \big\{\deg_G(w)\bigmid w\in N_G(v') \big\}.
    \end{equation}
    It is easy to express $\CQ_2$ by a $\GFOC$-formula. It turns out
    that, again, it is
    also possible to express $\CQ_2$ by an $\MFOC$-formula. Realising
    this was a key step for us towards proving that the two
    logics have the same expressive power.

    The idea is that we hash the sets in \eqref{eq:5} to numbers of
    size polynomial in the size of the input graph, then pass the hash
    value across the modal quantification, and compare the hash
    values. Then if the hashing produces no collisions, we will get
    the correct answer. Of course we need to carry out the
    collision-free hashing within the logic and without randomness,
    but this turns out to be not too difficult (we refer the reader to
    Section~\ref{sec:proof-logic}).

    \item Now consider the query $\CQ_3$ which is the same as $\CQ_2$,
    except that we replace the equality in \eqref{eq:5} by
    $\subseteq$. That is, $\CQ_3$ selects all nodes $x$ that have a
    neighbour $x'$ such that sets of degrees of the neighbours of $x$
    is a subset of the degrees of the neighbours of $x'$.

    Unfortunately, the same hashing construction no longer works,
    because if the two sets in \eqref{eq:5} are distinct they get
    different hash values, and since it is in the nature of hash
    functions that they are hard to invert, we cannot easily check if
    one set is a subset of the other. Yet we find a way around this
    (again, we refer the
    reader to Section~\ref{sec:proof-logic}).
    \end{enumerate}
\end{example}

Fully developing these ideas, we will be able to prove our main
logical result.

\begin{theorem}\label{thm:main-logic-intro}
  $\MFOC$ and $\GFOC$ have the same expressivity.
\end{theorem}

Since a unary query is expressible in $\MFOC$ if and only if it is
computable by a family 1-GNNs of polynomial size and bounded depth,
and similarly for $\GFOC$ and 2-GNNs, Theorem~\ref{thm:main-gnn}(3)
follows.

Let us close this introduction with a brief discussion of related work
on modal and guarded counting logics. Modal and guarded counting
logics have mostly been studied as fragments of $\LC^2$; the two
variable-fragment of $\LC$ (e.g.\
\cite{GradelOR97,Pratt-Hartmann07,KieronskiPT18}). Our logic $\MC$ is
equivalent to a logic known as \emph{graded modal logic}
\cite{Fine72,Tobies01,Otto19}. There is also related work on
description logics with counting (e.g.\
\cite{BaaderB19,HollunderB91}). However, all this work differs from
ours in at least two important ways: (1) the main question studied in
these papers is the decidability and complexity of the satisfiability
problem, and (2) the logics are interpreted over directed
graphs. For the logics as we define them here, the equivalence
results between modal and guarded fragments only hold over undirected
graphs. The modal and guarded fragments of $\FOC$ were introduced in
\cite{Grohe23} in the context of GNNs. 

The rest of the paper is organised as follows. After giving the
necessary preliminaries in Section~\ref{sec:prel}, in
Section~\ref{sec:counting-logics} we introduce the counting extensions
$\LC$ and $\FOC$ of first-order logic and their respective modal and
guarded fragments. The main result we prove is
\cref{thm:main-logic-intro}. In
Section~\ref{sec:gnn} we introduce graph neural networks and review
their logical characterisations. Assertions (2) and (3) of
\cref{thm:main-gnn} then follow (as Corollaries~\ref{cor:nonuniform}
and \ref{cor:bounded-nonuniform}). In Section~5, we discuss uniform
GNN-expressivity and prove \cref{thm:main-gnn}(1) (as
Theorem~\ref{theo:unisep}).

\emph{For all details omitted in the main part of this paper, we
refer the reader to the appendix.}

\section{Preliminaries}
\label{sec:prel}

By $\Nat,\PNat,\Real$ we denote the sets of nonnegative integers,
positive integers, and reals, respectively. For a set $S$, we denote
the power set of $S$ by $2^S$ and the set of all finite multisets with
elements from $S$ by $\multiset{S}{*}$. For $n\in\Nat$, we use the standard
notation $[n]\coloneqq\{i\in\Nat\mid 1\le i\le n\}$ as well as the variant
$\set{n}\coloneqq\{i\in\Nat\mid 0\le i<n\}$. We denote the $i$th bit
of the binary representation of $n$ by $\Bit(i,n)$; so
$n=\sum_{i\in\Nat}\Bit(i,n)2^i$. For $r\in\Real$ we let
$\sign(r)\coloneqq -1$ if $r<0$, $\sign(r)\coloneqq0$ if $r=0$, and
$\sign(r)\coloneqq1$ if $r>0$. We denote vectors of reals, but also
tuples of variables or tuples of vertices of a graph by lower-case
boldface letters. In linear-algebra contexts, we think of
vectors as column vectors, but for convenience we usually write all
vectors as row vectors.

\subsection{Feedforward Neural Networks}
We briefly review standard 
fully-connected feed-forward neural networks (FNNs), a.k.a.~multilayer
perceptrons (MLPs).
An \emph{FNN layer} of input dimension $m$ and output dimension $n$
computes a function from $\Real^m$ to $\Real^n$ of the form
\[
  \sigma\Big(A\vec x+\vec b\Big)
\]
for a \emph{weight matrix} $A\in \Real^{n\times m}$, a
\emph{bias vector} $\vec b\in\Real^{n}$, and an \emph{activation
  function} $\sigma:\Real\to\Real$ that is applied coordinatewise to
the vector $A\vec x+\vec b$.

An \emph{FNN} is a tuple $\big(L^{(1)},\ldots,L^{(d)})\big)$ of FNN layers, where the output dimension $n^{(i)}$ of $L^{(i)}$
matches the input dimension $m^{(i+1)}$ of $L^{(i+1)}$. Then the FNN computes a function from
$\Real^{m^{(1)}}$ to $\Real^{n^{(d)}}$, the composition
    $f^{(d)}\circ\cdots\circ f^{(1)}$ of the functions $f^{(i)}$ computed by
    the $d$ layers. 

    Typical activation functions used in practice are the \emph{logistic
  function} $\sig(x)\coloneqq(1+e^{-x})^{-1}$ (a.k.a. \emph{sigmoid
  function}), the \emph{hyperbolic tangent}
$\tanh(x)$, and the
\emph{rectified linear unit} $\relu(x)\coloneqq\max\{x,0\}$. 
In this paper (as in most other theoretical papers on graph neural
networks), \emph{unless explicitly stated otherwise we assume that
  FNNs only use $\relu$ and the identity function as activation
  functions, and we assume that all weights are rational numbers.}
These assumptions can be relaxed, but they will be convenient. One
needs to be careful with such assumptions (see
Example~\ref{exa:act}), but they are not the core issue studied
in this paper.

In a machine learning setting, we only fix the shape or architecture
of the neural network and learn the parameters (that is, the weights and biases). However, our focus in this paper is not on learning but on
the \emph{expressivity}, that is, on the question which functions can be
computed by FNNs. A fundamental
expressivity result for FNNs is the \emph{universal approximation
  theorem}: every continuous function $f:K\to\Real^n$ defined on a
compact domain $K\subseteq \Real^m$ can be approximated to any given
additive error by an FNN with just two layers.

\subsection{Graphs and Signals}
We use a standard notation for graphs. Graphs are always finite,
simple (no loops or parallel edges). and undirected. The \emph{order}
of a graph $G$, denoted by $|G|$, is the number of its vertices. The set
of neighbours of a vertex $v$ in a graph $G$ is denoted by $N_G(v)$,
and the \emph{degree} of $v$ is $\deg_G(v)\coloneqq|N_G(v)|$.  We
often consider vertex-labelled graphs, which formally we view as
graphs expanded by unary relations. For a set $\Lambda$ of unary
relation symbols, a \emph{$\Lambda$-labelled graph} is a
$\{E\}\cup\Lambda$-structure $G$ where the binary relation $E(G)$ is
symmetric and anti-reflexive. If $\Lambda=\{P_1,\ldots,P_\ell\}$, we
also call $G$ an \emph{$\ell$-labelled graph}, and we denote the
class of all $\ell$-labelled graphs by $\CG_\ell$. 

When serving as data for graph neural networks, the vertices of graphs
usually have real-valued features, which we call \emph{graph signals}.
An \emph{$\ell$-dimensional signal} on a graph $G$ is a function
$\Cf:V(G)\to\Real^\ell$. We denote the class of all $\ell$-dimensional
signals on $G$ by $\CS_\ell(G)$ and the class of all pairs $(G,\Cf)$,
where $\Cf$ is an $\ell$-dimensional signal on $G$, by $\CGS_\ell$.
An $\ell$-dimensional signal is \emph{Boolean} if its range is
contained in $\{0,1\}^\ell$. By $\CS^\bool_\ell(G)$ and $\CGS^\bool_\ell$ we
denote the restrictions of the two classes to Boolean signals. Observe
that there is a one-to-one correspondence between $\CG_\ell$ and
$\CGS^\bool_\ell$: the $\ell$-labelled graph $G\in\CG_\ell$
corresponds to $((V(G),E(G)),\Cb)\in \CGS^\bool_\ell$ for
$\Cb:V(G)\to\{0,1\}^\ell$ defined by $\big(\Cb(v)\big)_i=1\iff v\in
P_i(G))$. In the following, we will think of an $\ell$-labelled
$G\in\CG_\ell$ and the corresponding $((V(G),E(G)),\Cb)\in
\CGS^\bool_\ell$ as the same object and hence of $\CG_\ell$ and
$\CGS^\bool_\ell$ as the same class.

Isomorphisms between pairs $(G,\Cf),(H,\Cg)\in\CGS_\ell$ are required to
preserve the signals. We call a mapping $T:\CGS_\ell\to\CGS_m$ a \emph{signal
  transformation} if for all $(G,\Cf)\in\CGS_\ell$ we have
$T(G,\Cf)=(G,\Cf')$ for some $\Cf'\in\CS_m(G)$. Such a signal
transformation $T$
is \emph{equivariant} if for all isomorphic
$(G,\Cf),(H,\Cg)\in\CGS_\ell$, every isomorphisms
$h$ from $(G,\Cf)$ to $(H,\Cg)$ is also an isomorphism from $T(G,\Cf)$
to $T(H,\Cg)$.

\subsection{Logics and Queries}

We study the expressivity of logics and graph neural networks. Graph
neural networks operate on graphs and signals, and they compute signal
transformations. The logics we study operate on labelled graphs, and
they compute queries.  A \emph{$k$-ary query} on the class of
$\ell$-labelled graphs is a mapping $Q$ that associates with every
graph $G\in\CG_\ell$ a set ${\CQ}(G)\subseteq V(G)^k$ subject to the
following invariance condition: for all isomorphic graphs
$G,H\in\CG_\ell$ and isomorphisms $h$ from $G$ to $H$ it holds that
$h({\CQ}(G))={\CQ}(H)$. Observe that the correspondence between
$\CG_\ell$ and $\CGS_\ell^\bool$ extends to a correspondence between
unary queries and equivariant signal transformations
$\CGS_\ell^\bool\to\CGS_1^\bool$. More generally, equivariant signal
transformations can be viewed as generalisations of unary queries to
the world of graphs with real valued instead of Boolean
features. Similarly, $0$-ary (a.k.a.~Boolean) queries correspond to
invariant mappings $\CGS_\ell^\bool\to\{0,1\}$. If we wanted to lift
the correspondence to $k$-ary queries for $k\ge 2$, we would have to
consider signals on $k$-tuples of vertices.

When comparing the expressivity of logics, we compare the class of queries
definable in these logics. The logics we study here are ``modal'' and
most naturally define unary queries. Therefore, we restrict our
attention to unary queries. For logics $\LL,\LL'$, we say that $\LL$
is \emph{at most as expressive} as $\LL'$ (we write $\LL\preceq\LL'$)
if every unary query expressible in $\LL$ is also expressible in
$\LL'$. We say that $\LL$ is \emph{less expressive} than $\LL'$ (we
write $\LL\prec\LL'$) if $\LL\preceq\LL'$ but
$\LL'\not\preceq\LL$. We say that $\LL$ and $\LL'$ are \emph{equally
  expressive} (we write $\LL\equiv\LL'$) if $\LL\preceq \LL'$ and
$\LL'\preceq\LL$.

\section{First-order Logic with Counting}
\label{sec:counting-logics}
$\LC$ is the
extension of first-order logic $\FO$ by \emph{counting
  quantifiers} $\exists^{\ge p}$. That is, $\LC$-formulas are formed
from \emph{atomic formulas} of the form $x=y$, $E(x,y)$, $P_i(x)$ with the
usual Boolean connectives, and the new counting quantifiers. Standard
existential and universal quantifiers can easily be expressed using
the counting quantifiers ($\exists x$ as $\exists^{\ge 1}x$ and
$\forall x$ as $\neg\exists^{\ge 1}x\neg$). For every $k\ge 1$,
$\LC^k$ is the
$k$-variable fragment of $\LC$. While it is easy to see that every
formula of $\LC$ is equivalent to an $\FO$-formula, this is no longer
the case for the $\LC^k$, because simulating a quantifier
$\exists^{\ge n}$ in $\FO$ requires $n$ distinct
variables.

The logic $\LC$ treats numbers (cardinalities) as constants: the $n$
in a quantifier $\exists^{\ge n}$ is ``hardwired'' into the formula
regardless of the input graph. $\FOC$ is a more expressive counting
extension of $\FO$ in which numbers are treated as first class
citizens that can be compared, quantified, and manipulated
algebraically. The logic $\FOC$ has \emph{formulas} and \emph{terms}. Terms, taking values in
$\Nat$, are either dedicated \emph{number variables} ranging over $\Nat$, or
the constants $0,1,\ord$ ($\ord$ is always interpreted by the order of the
input graph), or \emph{counting terms} of the form
\begin{equation}
  \label{eq:counting-term}
    \#(x_1,\ldots,x_k,y_1<\theta_1,\ldots,y_\ell<\theta_\ell).\psi, 
\end{equation}
where the $x_i$ are \emph{vertex variables} ranging over the vertices
of a graph, the $y_i$ are number
variables, $\psi$ is a formula, and the $\theta_i$ are
terms. 
Furthermore, terms can be combined using addition and
multiplication. Formulas
are formed from atomic formulas $R(x_1,\ldots,x_k)$, $x=x'$ and term
comparisons $\theta\le\theta'$ using Boolean combinations. We do not
need quantifiers, because we can simulate them by counting terms. 

To define the semantics, we
think of formulas and terms as being interpreted in the 2-sorted expansion
$G\dot\cup(\Nat,+,\cdot,0,1)$ of a graph $G$. Vertex variables range
over $V(G)$ and number variables range over $\Nat$. We inductively
define a Boolean value for each formula and a numerical value in $\Nat$
for each term. This is straightforward, except for the case of
counting terms. The value of a counting term \eqref{eq:counting-term} is the
number of tuples $(v_1,\ldots,v_k,i_1,\ldots,i_\ell)\in
V(G)^k\times\Nat^\ell$ such that for all $j$, $i_j$ is smaller than
the value of the term $\theta_j$ and $\psi$ holds under the assignment
$x_i\mapsto v_i,y_j\mapsto i_j$.

An \emph{$\FOC$-expression} is either a term or a formula. We denote
terms by $\theta, \eta$ and variants such as
$\theta',\eta_1$, formulas by $\phi,\psi,\chi$ and variants, an
expressions by $\xi$ and variants.
We denote vertex variables by $x$ and number variables by $y$. We use $z$ for both types of variables. 
For a tuple $\vec z=(z_1,\ldots,z_k)$ of variables and a graph $G$, 
we let
$G^{\vec z}$ be the set of all tuples $(s_1,\ldots,s_k)\in
(G\cup\Nat)^k$ such that $s_i\in V(G)$ if $z_i$ is a vertex variable
and $s_i\in\Nat$ if $z_i$ is a number variable. 
We define the free
variables of an expression in the natural way, where a counting term
\eqref{eq:counting-term} binds the variables $x_1,\ldots,x_k,y_1,\ldots,y_\ell$.
For an
$\FOC$-expression $\xi$, we write $\xi(z_1,\ldots,z_k)$ or $\xi(\vec
z)$ to
indicate that the free variables of $\xi$ are among
$z_1,\ldots,z_k$. By $\xi^G(s_1,\ldots,s_k)$ we denote the
value of $\xi$ if the variables $z_i$ are interpreted by the
respective $s_i$. If $\xi$ is a formula, we also write
$G\models\xi(s_1,\ldots,s_k)$ instead of
$\xi^G(s_1,\ldots,s_k)=1$.
An expression is \emph{closed} if it has no free variables. For a
closed expression $\xi$, we denote the value by $\xi^G$, and if
it is a formula we write $G\models\xi$.

The reader may have noted that $\FOC$ does not have any
quantifiers. This is because existential and universal quantification
can easily be expressed using counting terms. For example,
$\exists x.\phi$ is equivalent to $1\le\#x.\phi$. We also have bounded
quantification over number variables. More generally, we can simulate
existential quantification
$\exists(x_1,\ldots,x_k,y_1<\theta_1,\ldots,y_\ell<\theta_\ell)$ over
tuples of vertex- and number variables. Once we have existential
quantification, we can simulate universal quantification using
existential quantification and negation. We use all the standard
Boolean connectives, and in addition we also use (inequalities)
$<,=,\ge,>$ over terms as they can all be expressed using
$\le$. Counting operators bind stronger than inequalities, which bind
stronger than negation, which binds stronger than other Boolean
connectives. We omit parentheses accordingly.

For $k\ge 1$, the \emph{$k$-variable fragment $\FOC[k]$} of $\FOC$
consists of all formulas with at most $k$ vertex variables. The number
of number variables in an $\FOC[k]$-formula is not restricted.

It is
easy to see that
\begin{equation*}
  \label{eq:2}
    \LC^k\prec\FOC[k]. 
\end{equation*}
To prove this, we
note that $\LC^k\preceq\FOC[k]$, because $\exists^{\ge n}x\phi$ can be
expressed as
\[
  \underbrace{1+\ldots+1}_{n\text{ times}}\le\#x.\phi.
\]
The inclusion is strict, in fact, we even have $\FOC[1]\not\preceq
\LC$, because the $\FOC[1]$-formula $\exists y.2\cdot y = \# x.x=x$
expresses that a graph has an even number of vertices, and it is not
hard to see (and well-known) that there is no $\FO$-formula and hence
no $\LC$-formula expressing even cardinality.

\subsection{The Modal and Guarded Fragments}

We will introduce two closely related fragments, a \emph{modal} and a
\emph{guarded} fragment, of the 2-variable logics $\LC^2$
and $\FOC[2]$. We assume that $\LC^2$ and $\FOC[2]$ only use the
vertex variables $x_1,x_2$ and, in the case of $\FOC[2]$, possibly
additional number variables. We use $x,x'$ to refer to either of these
variables, always stipulating that $x'\neq x$ (that is, either $x=x_1$
and $x'=x_2$ or $x=x_2$ and $x'=x_1$). For a formula $\phi(x,\vec
y)$ in $\LC^2$ or $\FOC[2]$,
we let $\phi(x',\vec y)$ be the formula obtained by simultaneously
replacing all occurrences of $x$ in $\phi(x,\vec y)$ by $x'$ and all
occurrences of $x'$ by $x$. Then the resulting formula is also in $\LC^2$ or
$\FOC[2]$, respectively.

A $\LC^2$-formula is \emph{modal} if for all subformulas 
$\exists^{\ge n}x'.\phi$, the formula $\phi$ is of the form
$\big(E(x,x')\wedge\psi(x')\big)$. Remember that $\psi(x')$ means
that only the variable $x'$, but not $x$, may occur freely in
$\psi$. 
The \emph{modal fragment} $\MC$ of $\LC$ consist of all modal $\LC^2$-formulas.

A $\LC^2$-formula is \emph{guarded} if for all subformulas 
$\exists^{\ge n}x'.\phi$, the formula $\phi$ is of the form
$\big(E(x,x')\wedge\psi(x,x')\big)$. 
The guarded fragment $\GC$ consist of all guarded $\LC^2$-formulas.

\begin{restatable}{proposition}{propmcvsgc}\label{prop:MCvsGC}
  \hspace{1cm}$
    \MC\equiv\GC.
  $
\end{restatable}

The simple proof is based on the
observation that all formulas $\psi(x,x')$ in \MC\ or \GC\ with two
free variables are just Boolean combinations of formulas with one free
variable and atoms $E(x,x')$, $x=x'$. 

The definitions for the modal and guarded fragments of $\FOC$ are
similar, but let us be a bit more formal here. The sets of
$\MFOC$-formulas and $\MFOC$-terms (in the language of
$\Lambda$-labelled graphs) are defined inductively as follows:
\begin{eroman}
\item all number variables and $0,1$ and $\ord$ are $\MFOC$-terms;
\item
  $\theta+\theta'$ and $\theta\cdot\theta'$ are $\MFOC$-terms,  for all $\MFOC$-terms $\theta,\theta'$;
\item $\theta\le\theta'$ is an $\MFOC$-formula, for all $\MFOC$-terms
  $\theta,\theta'$;
\item $x_i=x_j$, $E(x_i,x_j)$, $P(x_i)$ are $\MFOC$-formulas, for
  $i,j\in[2]$ and $P\in\Lambda$;
\item 
  $\#(y_1<\theta_1,\ldots,y_k<\theta_k).\psi$
  is an \MFOC-term, for all number variables 
  $y_1,\ldots,y_k$, all \MFOC-terms $\theta_1,\ldots,\theta_k$, and 
  all \MFOC-formulas $\psi$. 
\item
  $\#(x_{3-i},y_1<\theta_1,\ldots,y_k<\theta_k).\big(E(x_i,x_{3-i})\wedge\psi\big)$
  is an \MFOC-term, for all $i\in[2]$, all number variables
  $y_1,\ldots,y_k$, all \MFOC-terms $\theta_1,\ldots,\theta_k$, and
  all \MFOC-formulas $\psi$ such that $x_i$ does not occur freely in
  $\psi$.
\end{eroman}
The sets of
$\GFOC$-formulas and $\GFOC$-terms are defined inductively using rules
(i)--(v) and the rule (vi') obtained from (vi) by dropping the requirement
that $x_i$ does not occur freely in $\psi$. That is, in (vi'), $\psi$ may be
an arbitrary \GFOC-formula.

\begin{example}
  Recall Example~\ref{exa:intro}. The formula $\phi_1(x)$ in
  \eqref{eq:6} is guarded (that is, a $\GFOC$-formula). The term
  $\logic{deg}(x)$ and the formula in \eqref{eq:8} are modal.
\end{example}

Note that every formula of $\MFOC$ and $\GFOC$ either is purely
arithmetical, using no vertex variables at all and hence only
accessing the input graph by its order, or it has at least one \emph{free}
vertex variable. We are mainly interested in unary queries expressible
in the logics and thus in formulas with exactly one free vertex
variable and no free number variables. Note that an \MFOC-formula with only one free vertex
variable cannot contain a term formed by using rule (v) for a formula
$\psi$ with two free vertex variables, because in $\MFOC$, a formula
with two free vertex variables can never appear within a counting term
(necessarily of type (vi)) binding a free vertex variable.

Recall Theorem~\ref{thm:main-logic-intro} stating that 
\[
  \MFOC\equiv\GFOC.
\]
Before we prove the theorem in Section~\ref{sec:proof-logic}, we
continue with a few remarks on nonuniform expressivity.

\subsection{Nonuniform Expressivity}

In complexity theory, besides
the ``uniform computability'' by models like Turing machines, it is
common to also study ``nonuniform computability'', most often by
nonuniform families of circuits. Similarly, in the literature on graph
neural networks it is common to consider a nonuniform notion of
expressivity.

To capture non-uniformity in descriptive complexity theory, Immerman
(see \cite{Immerman99}) introduced a notion of nonuniform definability
that is based on so-called built-in relations. For $\FOC$, a convenient way of introducing
non-uniformity via built-in numerical relations was proposed in
\cite{Grohe23, KuskeS17}.
A \emph{numerical relation} is a relation 
$R\subseteq\Nat^k$ for some $k\in\Nat$. We extend the logic $\FOC$ by
new atomic formulas $R(y_1,\ldots,y_k)$ for all $k$-ary
numerical relations $R$ and number variables $y_1,\ldots,y_k$, with
the obvious semantics. (Of course only finitely many numerical
relations can appear in one formula.) By $\FOC\nuni$ we denote the extension of $\FOC$
to formulas using arbitrary numerical relations. We extend
the notation to fragments of $\FOC$, so we also have 
$\MFOC\nuni$ and $\GFOC\nuni$.


It is a well-known fact that on ordered graphs, $\FOC\nuni$ captures
the complexity class non-uniform $\TC^0$~\cite{BarringtonIS90}.

The proof of Theorem~\ref{thm:main-logic-intro} goes through in the
nonuniform setting without any changes. So as a corollary (of the
proof) we obtain the following.

\begin{corollary}\label{cor:main-logic}
  \hspace{1cm}$
    \MFOC\nuni\equiv\GFOC\nuni.
 $
\end{corollary}

Numerical built-in relations are tied to the two-sorted framework of
$\FOC$ with its number variables and arithmetic. There is a simpler
and more powerful, but in most cases too powerful notion of nonuniform
definability that applies to all logics. We say that a query $\CQ$ is
\emph{non-uniformly expressible} in $\LL$ if there is a family
$(\phi_n)_{n\in\Nat}$ of $\LL$-formulas such that for each $n$ the
formula $\phi_n$ expresses ${\CQ}$ on graphs of order $n$. This notion is
too powerful in the sense that if $\LL$ is at least as expressive as
first-order logic, then every query is non-uniformly expressible in
$\LL$. This is an easy
consequence of the fact that every finite graph can be described up to
isomorphism by a formula of first-order logic. For our counting
logics, we obtain the following simple Lemma~\ref{lem:nuex}, which may be regarded
as folklore. Assertion (1) essentially goes back to \cite{ImmermanL90}
and assertion (2) to \cite{Otto97}. For background on the Color
Refinement algorithm and Weisfeiler-Leman algorithm and their relation
to counting logics as well as graph neural networks, we refer the
reader to \cite{Grohe21}. We say that a unary query ${\CQ}$ is
\emph{invariant under Colour Refinement} if for all graphs
$G,H$ and vertices $v\in V(G)$, $w\in V(H)$ such that the algorithm
assigns the same colour to $v,w$ when run on $G,H$, respectively, we
have $v\in {\CQ}(G)\Leftrightarrow w\in {\CQ}(H)$. 

\begin{lemma}\label{lem:nuex}
  \begin{enumerate}
  \item A unary query ${\CQ}$ is non-uniformly expressible in $\MC$
    (and hence in $\GC$) if and only if it is invariant under Colour
    Refinement.
  \item Every unary query expressible in $\GFOC\nuni$ is 
    non-uniformly expressible in $\GC$ (and hence in $\MC$).
  \end{enumerate}
\end{lemma}

Note that assertion (2) follows from (1) by observing that every unary
query expressible in $\GFOC\nuni$ in invariant under Colour
Refinement.

\subsection{Proof of Theorem~\ref{thm:main-logic-intro}}
\label{sec:proof-logic}

To prove the theorem, \emph{we only consider graphs $G$ of order at
  least $n_0$}, where $n_0\ge 2$ is a constant determined later (in Lemma~\ref{lem:primes}). In particular, when we say that two formulas
are equivalent, this means that they are equivalent in graphs of order
at least $n_0$.

This assumption is justified by the fact that on graphs of order less than $n_0$, the logics $\MFOC$, $\GFOC$ have
the same expressivity as $\MC$. This follows from
Lemma~\ref{lem:nuex}(2).

In the proof, we informally distinguish
between \emph{small numbers} in $n^{O(1)}$ and \emph{large numbers} in
$2^{n^{O(1)}}$, where $n$ is the order of the input graph. We denote
small numbers by lower
case letters, typically $k,\ell,m,n,p$, and large numbers by uppercase letters, typically
$M,N$. Thus in space polynomial in $n$ we can represent small and
large numbers as well as sets of small numbers, but not sets of large
numbers.  

We first simplify our formulas.
We use the following lemma
\cite[Lemma~3.2]{Grohe23a}.


\begin{lemma}[\cite{Grohe23a}]\label{lem:termbound}
  For every $\FOC$-term $\theta(x_1,\ldots,x_k,y_1,\ldots,y_k)$
  there is a polynomial $\Fp_\theta(X)$ such that for all graphs $G$, all
  $v_1,\ldots,v_k\in V(G)$, and all $n_1,\ldots,n_\ell\in\Nat$ it
  holds that
  \[
    \theta^G(v_1,\ldots,v_k,n_1,\ldots,n_\ell)\le
    \Fp_\theta\Big(\max\big(\big\{|G|\big\}\cup\big\{n_i\bigmid i\in[\ell]\big\}\big)\Big).
  \]
\end{lemma}

Let $\phi$ be an $\FOC$-formula. Then
every number variable $y$ of $\phi$ except for the free number variables is \emph{introduced} in a counting term
$\#(x_1,\ldots,x_k,y_1<\theta_1,\ldots,y_\ell<\theta_\ell).\psi$. If
$y=y_i$, then it is bound by the term $\theta_i$, which we call the
\emph{bounding term} for $y$. We assume that all free number variables
$y$ of $\phi$ have a \emph{degree} $\deg(y)$. We inductively
define the \emph{degree} $\deg_\phi(y)$ of a number variable $y$ in
$\phi$ as follows. 
If $y$ is a free variable of $\phi$, then
$\deg_\phi(y)\coloneqq\deg(y)$. Otherwise, let 
$\theta=\theta(x_1,\ldots,x_k,y_1,\ldots,y_\ell)$ be the bounding term
for $y$, and let $\Fp_\theta(X)$ be the polynomial of
Lemma~\ref{lem:termbound}. Let $c\in\Nat$ be minimum such that
$\Fp_\theta(n)<n^c$ for all $n\ge 2$. Then if $\ell=0$, that is, the
bounding term $\theta$ has no free number variables, we let
$\deg_\phi(y)\coloneqq c$. Otherwise, we let $\deg_{\phi}(y)=cd$,
where $d\coloneqq\max\big\{\deg_{\phi}(y_i)\bigmid i\in[\ell]\}$.
Strictly speaking, we should have defined the degree of an occurrence
of a number variable $y$, because the same variable may be introduced
several times. But without loss of generality, we may assume that
every number variable is either free and never (re)introduced in a
counting term or introduced exactly once.

Observe that during the evaluation of $\phi$ in a graph of
order $n\ge2$ where each free variable $y$ of $\phi$ is assigned a
value smaller than $n^{\deg(y)}$, each number variable $y$ of $\phi$ can only take values less than $n^{\deg_\phi(y)}$. 

Let us call a formula $\phi$ \emph{simple} if
it satisfies the following two conditions:
\begin{eroman}
\item For each bound number variable $y$ in $\phi$, the bounding term of $y$
  is $\ord^{\deg_\phi(y)}$;
\item counting terms
  $\theta\coloneqq\#(x_1,\ldots,x_k,y_1<\theta_1,\ldots,y_\ell<\theta_\ell).\psi$
  only appear in subformulas $y=\theta$ for some number variable $y$
  that does not occur in $\theta$.
\end{eroman}
Note that (i) implies that all counting terms in $\phi$ are of the
form
$\#(x_1,\ldots,x_k,y_1<\ord^{d_1},\ldots,y_\ell<\ord^{d_i}).\psi$,
where $d_i=\deg_\phi(y_i)$. 
  
\begin{restatable}{lemma}{lemsimplenf}\label{lem:simplenf}
  Every $\FOC$-formula $\phi$ with no free number variables is
  equivalent to a simple $\FOC$-formula $\phi'$. Furthermore, if
  $\phi$ is in $\MFOC$ or $\GFOC$, then $\phi'$ is in $\MFOC$ or
  $\GFOC$, respectively, as well.
\end{restatable}

Next, we need to express some arithmetic on bit representations of
numbers. We think of a pair $\phi(y,\vec z),\theta(\vec z)$ consisting
of an $\FOC$-formula $\phi$ and an $\FOC$-term $\theta$ as representing a
number: for a graph $G$ and a tuple $\vec s\in G^{\vec z}$, we let
\[
  \num{\phi(\underline y,\vec z),\theta(\vec z)}^G(\vec s)\coloneqq
  \sum_{i\in\Nat,i<\theta^G(\vec s),G\models\phi(i,\vec s)}2^i.
\]
That is, $\num{\phi(\underline y,\vec z),\theta(\vec z)}^G(\vec s)$ is the
$N\in\Nat$ with $\Bit(i,N)=1\iff \big(i<\theta^G(\vec s)\text{ and
}G\models\phi(i,\vec s)\big).$ Note that we underline the variable $y$ in
$\num{\phi(\underline y,\vec z),\theta(\vec z)}$ to indicate which
variable we use to determine the bits of the number (otherwise it
could also be some variable in $\vec z$). The underlined variable does
not have to be the first in the list. 

An $\FOC$-expression is \emph{arithmetical} if it contains no vertex
variables. Note that all arithmetical $\FOC$-formulas are also
$\MFOC$-formulas and hence $\GFOC$-formulas.
The following lemma collects the
main facts that we need about the expressivity of arithmetical
operations in $\FOC$. Via the connection between $\FOC$ and the
circuit complexity class uniform $\TC^0$ \cite{BarringtonIS90}, the
results were mostly known in the 1990s, with the exception of division,
which was only established by Hesse in 2000 \cite{Hesse01,HesseAB02}. It is
difficult, though, to find proofs for these results in the
circuit-complexity literature. We refer the reader to \cite{Grohe23a}
for proof sketches.

\begin{lemma}[Folklore, \cite{BarringtonIS90,Hesse01,HesseAB02}]
  \begin{enumerate}
  \item
    There is an arithmetical $\FOC$-formula $\bit(y,y')$ such that for
    all graphs $G$ and all $i,n\in\Nat$ it holds that
    $G\models\bit(i,n)\iff\Bit(i,n)=1$.
  \item Let $\phi_1(y,\vec z), \phi_2(y,\vec z)$ be $\FOC$-formulas,
    and let $\theta_1(\vec z), \theta_2(\vec z)$ be $\FOC$-terms. Then
    there are $\FOC$-formulas $\logic{add}(y,\vec z)$,
    $\logic{sub}(y,\vec z)$ $\logic{mul}(y,\vec z)$,
    $\logic{div}(y,\vec z)$, and $\logic{leq}(\vec z)$ such that for
    all graphs $G$ and all $\vec s\in G^{\vec z}$ the following
    holds. For $i=1,2$, let
    $N_i\coloneqq \num{\phi_i(\underline y,\vec z),\theta_i(\vec
      z)}^G(\vec s)$. Then
    \[
      \begin{array}{r@{\,}c@{\,}l}
    \num{\logic{add}(\underline y,\vec
    z),\theta_1+\theta_2(\vec s)}^G(\vec s)&=&N_1+N_2,\\
    \num{\logic{sub}(\underline y,\vec
    z),\theta_1(\vec s)}^G(\vec s)&=&\max\{0,N_1-N_2\},\\
    \num{\logic{mul}(\underline y,\vec
    z),\theta_1+\theta_2(\vec s)}^G(\vec s)&=&N_1\cdot N_2,\\
\num{\logic{div}(\underline y,\vec
    z),\theta_1(\vec s)}^G(\vec
    s)&=&\floor{\frac{N_1}{N_2}}\quad\text{if $N_2\neq 0$},\\
G\models\logic{leq}(\vec
    s)&\iff& N_1\le N_2.
    \end{array}
  \]
  Furthermore, if the $\phi_i,\theta_i$ are arithmetical (modal,
  guarded) then
  $\logic{add},\logic{sub},\logic{mul},\logic{div},\logic{leq}$ are
  arithmetical (modal, guarded, respectively) as well.
  \end{enumerate}
\end{lemma}

\begin{corollary}\label{cor:mod}
  $\phi(y,\vec z)$ be an $\FOC$-formula,
  and let $\theta(\vec z)$ be an $\FOC$-term. Then there is an
  $\FOC$-formula $\logic{mod}(y,\hat y,\vec z)$ such that for all graphs
  $G$ and all $\vec s\in G^{\vec z}$, $n,\hat n\in\Nat$ we have
  \begin{align*}
    &G\models\logic{mod}(n,\hat n,\vec s)\\
    &\hspace{1cm}\iff 0\le n<\hat n\text{ and
    }n\equiv \num{\phi(\underline y,\vec z),\theta(\vec
      z)}^G(\vec s)\bmod{\hat n}.
  \end{align*}
   Furthermore, if the $\phi,\theta$ are arithmetical (modal,
  guarded) then
  $\logic{mod}$ is
  arithmetical (modal, guarded, respectively) as well.
\end{corollary}

The last ingredient for our proof of Theorem~\ref{thm:main-logic-intro} is a little
bit of number theory. A crucial idea of the proof is to hash sets of small numbers, or
equivalently, large numbers, to small ``signatures'', which will be
obtained by taking the large numbers modulo some small prime. This is
a variant of well-known perfect hash families based on primes.

\begin{restatable}{lemma}{lemhash}
  \label{lem:hash}
  Let $k,m\in\Nat$ and $\CM\subseteq\set{2^m}$. Furthermore, let $P\subseteq\Nat$
  be a set of primes of cardinality $|P|\ge km|\CM|^2$. Then
  \[
    \Pr_{p\in P}\big(\exists M,N\in\CM,M\neq N:\,M\equiv N\bmod p\big)<\frac{1}{k}.
  \]
  where the probability ranges over $p\in P$ chosen
  uniformly at random.
\end{restatable}

A second number theoretic fact we need is that we can find
sufficiently many small primes; this is a direct consequence of the prime number
theorem.

\begin{restatable}{lemma}{lemprimes}
  \label{lem:primes}
  There is an $n_0\ge 2$ such that for all $n\ge n_0$ there are at
  least $n$ primes $p\le 2n\ln n$.
\end{restatable}

Finally, we are ready to prove Theorem~\ref{thm:main-logic-intro}
(with some simple claims deferred to the appendix).

\begin{proof}[Proof of Theorem~\ref{thm:main-logic-intro}]
  By induction, we prove that for every $\GFOC$-formula
  $\phi(x,y_1,\ldots,y_k)$ and every assignment of a degree $\deg(y_i)$
  to each free number variable $y_i$ there is an $\MFOC$-formula
  $\hat\phi(x,y_1,\ldots,y_k)$ such that the following holds: for all
  graphs $G$ of order $n\coloneqq|G|$, all $v\in V(G)$, and all
  $a_1\in\set{n^{\deg(y_i)}},\ldots, a_k\in\set{n^{\deg(y_k)}}$ it holds
  that
  \[
    G\models\phi(v,a_1,\ldots,a_k)\iff
    G\models\hat{\phi}(v,a_1,\ldots,a_k).
  \]
  We may assume that $\phi$ is a
  simple formula. We let
  \[
    d\coloneqq\max\{\deg_\phi(y)\mid y\text{ number variable of
    }\phi\}.
  \]
  The only interesting step
  involves counting terms. Since $\phi$ is simple, this means that we
  have to consider a formula
  \begin{multline*}
    \phi(x,y_0,\ldots,y_k)=\Big(y_0=\#\big(x',y_{k+1}<\ord^{d_{k+1}},\ldots,y_{k+\ell}<\ord^{d_{k+\ell}}\big).\\
    \big(E(x,x')\wedge\psi(x,x',y_1,\ldots,y_{k+\ell})\big)\Big).
  \end{multline*}
  where for all $i\in[k+\ell]$ we let $d_i\coloneqq\deg_\phi(y_i)$.

  We need to understand the structure of $\psi$. Let
  us call maximal subformulas of $\psi$ with only one free vertex
  variable \emph{vertex formulas}. We distinguish between
  \emph{$x$-formulas}, where only $x$ occurs freely, and
  \emph{$x'$-formulas}, where only $x'$ occurs freely. The formula
  $\psi$ is formed from
  relational atoms $x=x',E(x,x')$, arithmetical formulas (that neither
  contain $x$ nor $x'$), and vertex formulas, using Boolean
  connectives, inequalities between terms, and counting terms of the
  form
  \begin{equation}
    \label{eq:3}
    \#(y_1'<\ord^{d_1'},\ldots,y_k'<\ord^{d_{\ell'}'}).\chi.
  \end{equation}
  We can
  argue that we do not need any relational atoms in $\psi$, because in
  $\phi$ we take the conjunction of $\psi$ with $E(x,x')$, which means
  that in $\psi$ the atom $E(x,x')$ can be set to``true'' and the atom
  $x=x'$ can be set to ``false''. Moreover, since the graph is
  undirected, we can also set $E(x',x)$ to ``true''. (Of course this
  only holds for atoms that are not in the scope of some counting
  term that binds $x$ or $x'$, that is, not contained in a vertex
  formula within $\psi$.) So we assume that $\psi$ is actually formed from
  arithmetical formulas and vertex formulas using Boolean connectives,
  inequalities between terms, and counting terms of the form
  \eqref{eq:3}.
  
  To turn $\phi$ into a modal formula, we need to eliminate the
  $x$-formulas in $\psi$. Let $\chi_1,\ldots,\chi_q$ be an enumeration
  of all $x$-formulas in $\psi$, where
  $\chi_i=\chi_i(x,y_{i,1},\ldots,y_{i,k_i})$. Here the $y_{i,j}$
  may be variables in $\{y_1,\ldots,y_{k+\ell}\}$, or they may be
  number variables $y_i'$ introduced by counting terms \eqref{eq:3}.

  Let $G$ be a graph of order $n$. Let $m\coloneqq n^d$. Since
  $d\ge\deg_\phi(y)$ for all number variables $y$ appearing in $\phi$, when evaluating $\phi$ in $G$, number
  variables only take values in $\set m$. Let $i\in[q]$. Then for
  every $v\in V(G)$, the formula $\chi_i$ defines a relation
  \[
    R_i(v)\coloneqq\big\{(a_1,\ldots,a_{k_i})\in\set{m}^{k_i}\bigmid
    G\models\chi_i(v,a_1,\ldots,a_{k_i})\big\}.
  \]
  With each tuple $(a_1,\ldots,a_{k_i})\in\set{m}^{k_i}$ we associate
  the (small) number
  \[
    \angles{a_1,\ldots,a_{k_i}}\coloneqq \sum_{j=1}^{k_i}a_j m^{j-1}\in\set{m^{k_i}}.
  \]
  Then we can encode $R_i(v)$ by the (large) number
  \[
    N_i(v)\coloneqq \sum_{(a_1,\ldots,a_{k_i})\in
      R_i(v)}2^{\angles{a_1,\ldots,a_{k_i}}}\in\set{2^{m^{k_i}}}.
  \]
  Let
  \[
    \CN(G)\coloneqq\big\{N_i(v)\bigmid i\in[q],v\in V(G)\}.
  \]
  Then $|\CN|\le qn$, and for $k_0\coloneqq\max\{k_i\mid i\in[q]\}$ we
  have $\CN\subseteq \set{2^{m^{k_0}}}$.

  We want to use small primes $p\in O(n^2)$ to hash the set $\CN$ to a
  set of small numbers. A prime $p$ is \emph{good for $G$} if for all
  distinct $N,N'\in\CN(G)$ it holds that $N\not\equiv N'\bmod p$.

  There is some $c\in\Nat$ that does not depend on $n$, but only on
  the formula $\phi$ and the parameters $d,k_0,q$ derived from $\phi$ such
  that
  \[
    n^{c}\ge 4q^2n^{dk_0+2}\ln(2q^2n^{dk_0+2})=4q^2m^{k_0}n^2\ln 2q^2m^{k_0}n^2.
  \]
  Let $P_n$ be the set of all primes less than or equal to $n^c$. Then
  by Lemma~\ref{lem:primes}, $|P_n|\ge 2q^2m^{k_0}n^2\ge2m^{k_0}|\CN|^2$.
  By Lemma~\ref{lem:hash} with
  $k\coloneqq2,m\coloneqq m^{k_0},\CM\coloneqq\CN$, more than half of the
  primes in $P_n$ are good.

  Now suppose $p\in P_n$ is a prime that is good for $G$.
  For every $i\in [q],v\in V$, we let $n_i(v,p)\in\set p$ such that
  $n_i(v,p)\equiv N_i(v)\bmod p$. Observe that 
  for all vertices $v,v'\in V(G)$, if $n_i(v,p)=n_i(v',p)$ then 
  $N_i(v)=N_i(v)$ and thus $R_i(v)=R_i(v')$. This means that for all
  $a_1,\ldots,a_{k_i}\in\set m$ it holds that
  \[
    G\models\chi_i(v,a_1,\ldots,a_{k_i})\iff
    G\models\chi_i(v',a_1,\ldots,a_{k_i}).
  \]

  \begin{claim}\label{claim:1}
    For every $i\in[q]$ there is a formula $\zeta_i(x,z,z_i)$ such
    that for all $v\in V(G)$ and $b\in\Nat$,
    \[
      G\models\zeta_i(v,p,b)\iff b=n_i(v,p).
    \]
    Here $z$ and $z_i$ are fresh number variables that do not occur in
    $\phi$.
%
    \end{claim}

  Crucially, the formulas $\zeta_i$ in Claim~\ref{claim:1} do not
  depend on the graph $G$, but only on the formula $\phi$. The same
  will hold for all formulas defined in the following. 

  For every $i\in [q]$, let
  \begin{multline*}
  \chi_i'(x', y_{i,1},\ldots,y_{i,k_i},z,z_i)\coloneqq\\
  \exists
    x\big(E(x',x)\wedge
    \zeta(x,z,z_i)\wedge\chi_i(x,y_{i,1},\ldots,y_{i,k_i})\big).
  \end{multline*}
  Note that $\chi_i'$ is an $\MFOC$-formula.

  \begin{claim}[resume]\label{claim:2}
    For all $(v,v')\in E(G)$ and $a_{1},\ldots,a_{k_i}\in\set{m}$
    it holds that
    \[
      G\models\chi_i(v,a_1,\ldots,a_{k_i})\iff G\models\chi_i'(v',
      a_1,\ldots,a_{k_i},p,n_i(v,p)).
    \]
%

  \end{claim}

  Now let $\psi'(x',y_1,\ldots,y_{k+\ell},z,z_1,\ldots,z_q)$ be the formula
  obtained from $\psi(x,x',y_1,\ldots,y_{k+\ell})$ by replacing, for each
  $i\in[q]$, the $x$-formula $\chi_i(x,y_{i,1},\ldots,y_{i,k_i})$ by
  the $x'$-formula $\chi_i'(x', y_{i,1},\ldots,y_{i,k_i},z,z_i)$. Then
  $\psi'$ is an \MFOC-formula.

   \begin{claim}[resume]\label{claim:3}
    For all $(v,v')\in E(G)$ and $a_{1}\in\set{n^{d_1}},\ldots,a_{k+\ell}\in\set{n^{d_{k+\ell}}}$
    it holds that
    \begin{align*}
      &G\models\psi(v,v',a_1,\ldots,a_{k+\ell})\\
      \iff&G\models\psi'(v',
      a_1,\ldots,a_{k+\ell},p,n_1(v,p),\ldots,n_q(v,p)).
    \end{align*}
%
  \end{claim}

  We let
  \begin{multline*}
    \phi'(x,y_0,\ldots,y_k,z)\coloneqq
    \exists
    (z_1<z,\ldots,z_q<z).\Bigg(\bigwedge_{i=1}^q\zeta_i(x,z,z_i)  \wedge\\
     y_0=\#\big(x',y_{k+1}<\ord^{d_{k+1}},\ldots,y_{k+\ell}<\ord^{d_{k+\ell}}\big).\\
    \big(E(x,x')\wedge\psi'(x',y_1,\ldots,y_{k+\ell},z,z_1,\ldots,z_q)\big)\Big),
  \end{multline*}
  Note that
  $\phi'$ is an \MFOC-formula.

  \begin{claim}[resume]\label{claim:4}
    For all $v\in V(G)$ and
    $a_{1}\in\set{n^{d_1}},\ldots,a_{k}\in\set{n^{d_k}}$,
    \begin{equation}
      \label{eq:4}
       G\models\phi(x,a_1,\ldots,a_k)\iff
      G\models\phi'(x,a_1,\ldots,a_k,p).
    \end{equation}
%
  \end{claim}

  In fact, the assertion of Claim~\ref{claim:4} holds for all primes
  $p$ that are good for $G$. Recall that more than half of the primes
  $p<n^c$ are good. Thus \eqref{eq:4} holds for more than half of the
  primes $p<n^c$. Let
  \[
    \logic{prime}(z)\coloneqq\forall(y<z,y'<z).\neg y\cdot y'=z,
  \]
  expressing that $z$ is a prime. Then we let
  \begin{multline*}
    \hat\phi(x,y_1,\ldots,y_k)\coloneqq\#
    z<\ord^c.\logic{prime}(z)\\
    <2\cdot\# z<\ord^c.\big(\logic{prime}(z)\wedge
    \phi'(x,y_1,\ldots,y_k,z)\big).
  \end{multline*}

    \begin{claim}[resume]\label{claim:5}
    For all $v\in V(G)$ and
    $a_{1}\in\set{n^{d_1}},\ldots,a_{k}\in\set{n^{d_k}}$,
    \begin{equation}
      \label{eq:4}
       G\models\phi(x,a_1,\ldots,a_k)\iff
      G\models\hat\phi(x,a_1,\ldots,a_k).\qedhere
    \end{equation}
  \end{claim}

\end{proof}

\section{Graph Neural Networks}
\label{sec:gnn}

Recall that we want to study the expressivity of \emph{GNNs with
  1-sided messages (1-GNNs)} and \emph{GNNs with 2-sided messages
  (2-GNNs)}. 
Both 1-GNNs and 2-GNNs consist of a
sequence of layers. 
A \emph{1-GNN layer} is a triple $\FL=(\msg,\agg,\comb)$
of functions: a \emph{message function}
$\msg:\Real^{p}\to\Real^{r}$, an \emph{aggregation function} $\agg:\multiset{\Real^r}{*}\to\Real^r$, and a \emph{combination function}
$\comb:\Real^{p+r}\to\Real^q$. We call $p$ the \emph{input dimension}
and $q$ the \emph{output dimension} of the layer. A \emph{2-GNN layer}
 is defined in exactly the same way, except that
 $\msg:\Real^{2p}\to\Real^{r}$.

 An $i$-GNN layer $\FL=(\msg,\agg,\comb)$ computes a signal
 transformation $T_\FL\colon
\CGS_p\to\CGS_q$. We let $T_{\FL}(G,\Cf)\coloneqq\big(G,
S_{\FL}(G,\Cf)\big)$, where 
 $S_{\FL}(G,\Cf)\in\CS_q(G)$ is the signal defined by
\begin{equation}
  \label{eq:20}
  S_{\FL}(G,\Cf)(v)\coloneqq \comb\Bigg(\Cf(v),\agg\Big(\Biglmulti
  \msg\big(\Cf(w)\big)\Bigmid w\in
  N_G(v)\Bigrmulti\Big)\Bigg)
\end{equation}
if $\FL$ is a 1-GNN-layer, and by
\begin{multline}
  \label{eq:21}
  S_{\FL}(G,\Cf)(v)\coloneqq\\
  \comb\Bigg(\Cf(v),\agg\Big(\Biglmulti
  \msg\big(\Cf(v),\Cf(w)\big)\Bigmid w\in
  N_G(v)\Bigrmulti\Big)\Bigg)
\end{multline}
if $\FL$ is a 2-GNN-layer. Note that the only difference between the
two is that in a 1-GNN the messages only depend on the state of
the sender, whereas in a 2-GNN they depend on the states of both
sender and recipient. 
 
The message function $\msg$ and combination function $\comb$ of a GNN
layer are computed by FNNs, and the parameters are typically learned
from data. To be precise, we need to add these neural networks rather
than the function they compute to the specification of a GNN
layer (that is, replace $\comb$ and $\msg$ by FNNs computing these
functions), but we find it more convenient here to just give the functions.
In principle, the aggregation function can be any function
from finite multisets of vectors to vectors, but the most common
aggregation functions used in practice are summation (\SUM),
arithmetic mean (\MEAN), and maximum (\MAX), applied
coordinatewise. We restrict our attention to these. We sometimes have
to aggregate over the empty multiset $\emptyset$, and we define
$\SUM(\emptyset)\coloneqq\MEAN(\emptyset)
\coloneqq\MAX(\emptyset)\coloneqq\vec0$.

For $i=1,2$, an \emph{$i$-GNN} is a tuple
$\FN=(\FL^{(1)},\ldots,\FL^{(d)})$ of $i$-GNN layers such that the output
dimension $q^{(i)}$ of $\FL^{(i)}$ matches the input dimension
$p^{(i+1)}$ of $\FL^{(i+1)}$. We call $p^{(1)}$ the
\emph{input dimension} of $\FN$ and $q^{(d)}$ the \emph{output dimension}.

An $i$-GNN $\FN=(\FL^{(1)},\ldots,\FL^{(d)})$ computes the signal
transformation
\[
  T_\FN\coloneqq T_{\FL^{(d)}}\circ T_{\FL^{(d-1)}}\circ\ldots\circ
  T_{\FL^{(1)}}\colon \CGS_{p^{(1)}}\to\CGS_{q^{(d)}}
\]
that is, the composition of the transformations of its layers. We also
define $S_{\FN}(G,\Cf)\in\CS_{q^{(d)}}(G)$ to be the signal such that $T_\FN(G,\Cf)\coloneqq\big(G, S_{\FN}(G,\Cf)\big)$.

We observe that we can actually simplify 1-GNNs by assuming that the
message function is always just the identity.

\begin{restatable}{lemma}{lemmsg}
  \label{lem:msg}
  For every 1-GNN $\FN=(\FL^{(1)},\ldots,\FL^{(d)})$ there is a 1-GNN
  $\tilde\FN=(\tilde\FL^{(0)},\tilde\FL^{(1)},\ldots,\tilde\FL^{(d)})$
  such that $T_{\FN}=T_{\tilde\FN}$ and the message function of each layer $\tilde\FL^{(i)}$ is
  the identity function on $\Real^{\tilde p^{(i)}}$.
  \end{restatable}

A similar claim for 2-GNNs does not seem to hold. If the message
function is linear, which in practice it often is, and the aggregation
function is $\MEAN$, then the message function can be pulled into the
combination function of the same layer and be replaced by
identity. 

\begin{restatable}{proposition}{proplinearmsg}
  Let $\FN$ be a 2-GNN such that on all layers of $\FN$, the message
  function is an affine linear function and the aggregation function
  is $\MEAN$. Then there is a 1-GNN $\tilde{\FN}$ such
  that $T_{\FN}=T_{\tilde\FN}$.
\end{restatable}

\subsection{Expressivity and Logical Characterisations}

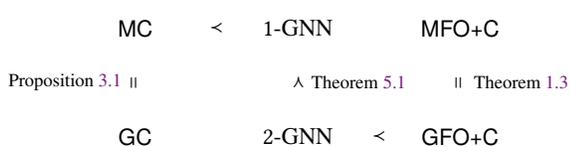
\begin{figure}
  \centering
  \begin{tikzpicture}[scale=0.9]
    \node (MC) at (0,1.6) {$\MC$}; 
    \node (GC) at (0,0) {$\GC$}; 
    \node (1G) at (2.4,1.6) {$1$-GNN}; 
    \node (1G) at (2.4,0) {$2$-GNN}; 
    \node (1G) at (4.8,1.6) {\MFOC}; 
    \node (1G) at (4.8,0) {\GFOC};
    \path (0,0.8) node[rotate=90] {$=$} node[left=2pt]{\footnotesize Proposition~\ref{prop:MCvsGC}}  (1.2,1.6) node {$\prec$} (2.4,0.8) node[rotate=-90] {$\prec$}
    node[right=2pt]{\footnotesize Theorem~\ref{theo:unisep}}
    (3.6,0) node {$\prec$} (4.8,0.8) node[rotate=90] {$=$} node[right=2pt]{\footnotesize Theorem~\ref{thm:main-logic-intro}};
  \end{tikzpicture}
  \caption{Uniform expressivity}\label{fig:uniform}
\end{figure}

GNNs compute signal transformations, that is,
equivariant functions on the vertices of a graph. To be able to compare
them to logics, we are interested in the queries computable by
GNNs. Let $\FN$ be a GNN of input dimension $\ell$ and output
dimension $1$, and $\CQ$ be a unary query on labelled graphs. We say
that $\FN$ \emph{(uniformly) computes} ${\CQ}$ if for all $(G,\Cb)\in\CGS_\ell^\bool$ (recall
that we do not distinguish between an $\ell$-labelled graph and the
corresponding graph with an $\ell$-dimensional
Boolean signal) and all $v\in V(G)$ we have 
\[
  S_{\FN}(G,\Cb)(v)
    \begin{cases}
      \ge \frac{3}{4}&\text{if }v\in \CQ(G,\Cb),\\
      \le\frac{1}{4}&\text{if }v\not\in\CQ(G,\Cb).
    \end{cases}
\]
Barcelo et al.~\cite{BarceloKM0RS20} proved that every
unary query expressible by an $\MC$-formula is expressible by
a 1-GNN with $\SUM$-aggregation and $\relu$ activation
(in the FNNs computing the message and combination functions). Grohe~\cite{Grohe23}
proved that every query expressible by a $2$-GNN is also expressible
in $\GFOC$. Essentially the same proof also yields that that every query expressible by a $1$-GNN is also expressible
in $\MFOC$. The results are illustrated in
Figure~\ref{fig:uniform}. All inequalities are
strict; for the inequality between $1$-GNN and $2$-GNN we prove this
in the next section.


As explained in the introduction, while uniform expressivity seems to be the most natural
notion of GNN-expressivity, much of the theoretical literature on
GNNs is concerned with a non-uniform notion where the GNN may depend
on the size of the input graph. We formalise non-uniform
GNN-expressivity as follows. Let
$\CN\coloneqq(\FN_n)_{n\in\Nat}$ be a family of
GNNs. Following~\cite{Grohe23}, we say that $\CN$
\emph{(non-uniformly) computes} a unary query ${\CQ}$ on $\ell$-labelled
graphs if for all $n\in\PNat$, all $(G,\Cb)\in\CGS_\ell^{\bool}$ of
order $|G|=n$, and
all $v\in V(G)$ it holds that
\[
  S_{\FN_n}(G,\Cb)(v)
    \begin{cases}
      \ge \frac{3}{4}&\text{if }v\in \CQ(G),\\
      \le\frac{1}{4}&\text{if }v\not\in\CQ(G).
    \end{cases}
  \]  
Since we consider GNNs with
rational weights here, we can take \emph{size} of a GNN to be the
bitsize of a representation of the GNN.  The \emph{depth} of a GNN is
the sum of the depths of the FNNs computing the message and
combination functions of the layers of the GNN. We say that a family $\CN\coloneqq(\FN_n)_{n\in\PNat}$ has
\emph{polynomial size} if there is a polynomial $\Fp$ such that the
size of $\FN_n$ is at most $\Fp(n)$.  We say that the
family has \emph{bounded depth} if there is a
constant $d$ such that for all $n$ the depth of $\FN_n$ is at most
$d$.

Let us first look at non-uniform expressivity without any restrictions
on the family $\CN$ of GNNs. We have the following logical
characterisation of non-uniform GNN expressivity; it is a direct consequence of the characterisation of
GNN-expressivity in terms of the Weisfeiler-Leman
algorithm~\cite{MorrisRFHLRG19,XuHLJ19}.

\begin{theorem}[\cite{MorrisRFHLRG19,XuHLJ19}]
  Let ${\CQ}$ be a unary query.
  \begin{enumerate}
  \item If $\CQ$ is (non-uniformly)
    computable by a family $\CN$ of 2-GNNs, then $\CQ$ is non-uniformly expressible in $\MC$.
  \item If $\CQ$ is non-uniformly expressible in $\MC$, then it is
    computable by a family of $1$-GNNs.
  \end{enumerate}
\end{theorem}

\begin{corollary}\label{cor:nonuniform}
  A unary query $\CQ$ is computable by a family of 1-GNNs if and only if it is computable by a family of 2-GNNs.
\end{corollary}

It has been proved in \cite{Grohe23} that non-uniform computability by
GNN families of polynomial size and bounded depth can be characterised
in terms of the modal and guarded fragments of $\FOC\nuni$, that is,
$\FOC$ with built-in relations.

\begin{theorem}[\cite{Grohe23}]
  Let ${\CQ}$ be a unary query.
  \begin{enumerate}
  \item ${\CQ}$ is expressible by a polynomial size bounded-depth family
    of 2-GNNs if and only if ${\CQ}$ is expressible in $\GFOC\nuni$.
  \item ${\CQ}$ is expressible by a polynomial size bounded-depth family
    of 1-GNNs if and only if ${\CQ}$ is expressible in $\MFOC\nuni$.
  \end{enumerate}
\end{theorem}

Actually, only assertion (1) of the theorem is proved in
\cite{Grohe23}, but the proof can easily be adapted to (2).

\begin{corollary}\label{cor:bounded-nonuniform}
  A unary query $\CQ$ is expressible by a polynomial-size
  bounded-depth family of 1-GNNs if and only if it is expressible by a polynomial-size
  bounded-depth family of 2-GNNs.
\end{corollary}

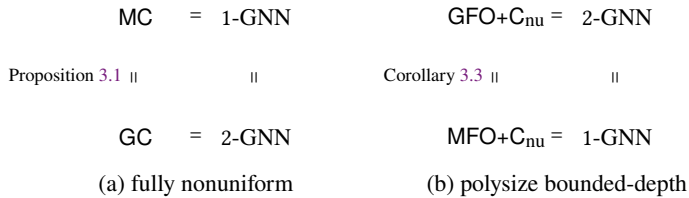
\begin{figure}
  \centering
  \begin{tikzpicture}[scale=0.8]
  \useasboundingbox (0.2,-0.8) rectangle (8.2,2);
    \begin{scope}
    \node (MC) at (0,0) {$\GC$}; 
    \node (GC) at (0,2) {$\MC$}; 
    \node (1G) at (2,0) {$2$-GNN}; 
    \node (1G) at (2,2) {$1$-GNN};
    \path (1,0) node {$=$} (1,2) node {$=$} (0,1) node[rotate=90]
    {$=$} node[left=2pt]{\footnotesize Proposition~\ref{prop:MCvsGC}}
    (2,1) node[rotate=90] {$=$} (1,-0.8) node {(a) fully
      nonuniform};
    \end{scope}

    \begin{scope}[xshift=6cm]
    \node (MC) at (0,0) {$\MFOC\nuni$}; 
    \node (GC) at (0,2) {$\GFOC\nuni$}; 
    \node (1G) at (2,0) {$1$-GNN}; 
    \node (1G) at (2,2) {$2$-GNN};
    \path (1,0) node {$=$} (1,2) node {$=$} (0,1) node[rotate=90]
    {$=$} node[left=2pt]{\footnotesize Corollary~\ref{cor:main-logic}} (2,1) node[rotate=90] {$=$} (1,-0.8) node {(b) polysize bounded-depth};
    \end{scope}
\end{tikzpicture}
  \caption{Nonuniform expressivity}\label{fig:nonuniform}
\end{figure}

Figure~\ref{fig:nonuniform} summarises the results in the non-uniform setting.

\section{Uniform Separation}
\label{sec:uniform}

In this section, we separate the expressivity of 1-GNNs and
2-GNNs. Interestingly, the separation only works for GNNs that only
use $\SUM$-aggregation. We will prove that 2-GNNs using only
\MEAN-aggregation or only \MAX-aggregation can be simulated by 1-GNNs
(using the same aggregation function).

\subsection{SUM Aggregation}

Let $\CQ_1$ be the query on unlabelled graphs that selects all
vertices that have a neighbour of larger degree. That is, for every
$G\in\CG_0$ we let
\[
  \CQ_1(G)=\{u\in V(G)\bigmid\exists v\in N_G(u):\;\deg_G(u)<\deg_G(v)\big\}.
\]
This is precisely the query considered in \cref{exa:intro}(1), and we
have already observed there that the query is expressible in $\MFOC$.

\begin{theorem}\label{theo:unisep}
  The query $\CQ_1$ is expressible by a 2-GNN with \SUM-aggregation,
  but not by a 1-GNN with \SUM-aggregation.
\end{theorem}

It is fairly obvious that $\CQ_1$ is expressible by a 2-layer
2-GNN. The first layer computes the degree of all vertices by summing the
constant message $1$ for all neighbours. On the second layer, the
message sent along an edge $(v,u)$ is $1$ if $\deg(u)<\deg(v)$
and $0$ otherwise. The combination function then just needs to check
if the sum of all messages is at least $1$.

The proof that $\CQ_1$ is not expressible by a 1-GNN requires some preparation.
We call a polynomial $\Fp(X,Y)$ in two variables \emph{nice} if it
is of the form
\begin{equation}
  \label{eq:1}
  \Fp(X,Y)=\sum_{i=0}^k a_i X^{\floor{i/2}} Y^{\ceil{i/2}} 
\end{equation}
with arbitrary real coefficients $a_i$.
 The \emph{leading coefficient} of $\Fp$ is $a_i$ for the maximum
 $i$ such that $a_i\neq0$, or $0$ if all $a_i$ are $0$.
We call a polynomial $\Fp(X,Y)$
\emph{co-nice} if $\Fp(Y,X)$ is nice.


\begin{lemma}
  \label{lem:nice-sign}
Let $\Fp(X,Y)$ be a nice or co-nice polynomial with leading
coefficient $a$. Then either $\Fp$ is constant and thus $\Fp(m,n)=a$
for all $m,n\in\Nat$, or there is an
$n_0\in\Nat$ such that for all $n\ge n_0$ we have
\begin{equation}
  \label{eq:12}
  \sign\big(\Fp(n-1,n)\big)=\sign\big(\Fp(n+1,n)\big)=\sign(a) 
\end{equation}
and
\begin{equation}
  \label{eq:13}
    |\Fp(n-1,n)|, |\Fp_1(n+1,n)|\ge \frac{|a|}{2}n.
\end{equation}

\end{lemma}

\begin{proof}
  There is nothing to prove if $\Fp$ is constant, so assume that
  it is not. Assume that $\Fp(X,Y)$ is co-nice. (The argument for nice
  $\Fp$ is similar and even slightly simpler.) Then
  \[
    \Fp(X,Y)=\sum_{i=0}^k
    a_i X^{\ceil{i/2}} Y^{\floor{i/2}}
  \]
  for some $k\ge 1$ and coefficients $a_i$ with $a_k=a\neq 0$. We
  write $\Fp$ as $\Fp=\Fp_1+\Fp_2$, where
  \[
    \Fp_1(X,Y)=a X^{\ceil{k/2}} Y^{\floor{k/2}},\quad
    \Fp_2(X,Y)=\sum_{i=0}^{k-1} a_i X^{\ceil{i/2}} Y^{\floor{i/2}}.
  \]
  It is easy to see that for sufficiently large $n_0$ and $m,n\ge n_0$ we have
  $|\Fp_2(m,n)|\le\frac{1}{4}|\Fp_1(m,n)|$, which implies
  \[
    \sign\big(\Fp(m,n)\big)=\sign\big(\Fp_1(m,n)\big)=\sign(a) 
  \]
  and hence \eqref{eq:12}. It also implies
  \[
    |\Fp(m,n)|\ge\frac{3}{4}\Fp_1(m,n)\ge\frac{3}{4}|a|m,
  \]
  where the last inequality holds because $k\ge 1$ and thus
  $\ceil{k/2}\ge 1$. Since for sufficiently large $n$ we have $n-1\ge
  \frac{2}{3}n$, \eqref{eq:13} follows.
\end{proof}

We say that a function $f:\Nat^2\to\Real$ is \emph{fast-converging}
to a polynomial $\Fp(X,Y)$ if for all $r\in\Real$ there is an
$n_0$ such that for all $n\ge n_0$ and $m\in\{n-1,n+1\}$ it
holds that 
\[
  \big|f(m,n)-\Fp(m,n) \big|\le \frac{1}{n^r}.
\]
Note that fast convergence only considers argument pairs $(n-1,n)$ or
$(n+1,n)$ for $n\in\PNat$.
We 
let $\FC(\Fp)$ be the class of all $f:\Nat^2\to\Real$ that are
fast-converging to $\Fp$. We let 
\[
  \FC\coloneqq\bigcup_{\Fp(X,Y)\text{
      nice}}\FC(\Fp)
\]
be the class of all functions that are
fast-converging to some nice polynomial. Similarly, for a co-nice polynomial $\Fp$ we denote the fast convergence of $f$ to $\Fp$ by 
$\FCco(f,\Fp)$, we let $\FCco(\Fp)$ be the class of all $f:\Nat^2\to\Real$ fast-converging to $\Fp$, and define $\FCco$ respectively.


A function $g:\Real^k\to\Real$ \emph{preserves fast convergence} if for
all $f_1,\ldots,f_k\in\FC$ it holds that $g(f_1,\ldots,f_k)\in\FC$
and for
all $f_1,\ldots,f_k\in\FCco$ it holds that
$g(f_1,\ldots,f_k)\in\FCco$.

Here $g(f_1,\ldots,f_k):\Nat^2\to\Real$ is the function defined by
\[
  g(f_1,\ldots,f_k)(x,y)\coloneqq g\big(f_1(x,y),\ldots,f_k(x,y)\big).
\]

\begin{restatable}{lemma}{lempreFC}
  \label{lem:preFC}
  \begin{enumerate}
  \item\label{it:pFC1} All constant functions $c:\Nat^2\to\Real$ are in
    $\FC\cap\FCco$.
  \item\label{it:pFC2} For $f:\Nat^2\to\Real$, define $f_1,f_2:\Nat^2\to\Real$ by
    $f_1(m,n)\coloneqq m\cdot f(n,m)$ and $f_2(m,n)\coloneqq n\cdot
    f(m,n)$. Then
    \begin{align*}
      f\in\FCco&\implies f_1\in\FC;\\
      f\in\FC&\implies f_2\in\FCco.
    \end{align*}
  \item\label{it:pFC3} All linear functions preserve fast convergence.
  \item\label{it:pFC4} All functions computed by FNNs that only use
    activation functions preserving fast convergence preserve fast convergence.
  \end{enumerate}
\end{restatable}

\begin{proof}
  All polynomials of degree $0$ (constants) are nice, which proves
  \ref{it:pFC1}.

  To prove~\ref{it:pFC2}, suppose that $f\in\FC(\Fp)$ for the polynomial $\Fp$
  in \eqref{eq:1}. Note that
  \[
    X\cdot\Fp(X,Y)=\sum_{i=0}^k a_i X^{\floor{i/2}+1}
    Y^{\ceil{i/2}}=\sum_{i=0}^k a_i X^{\ceil{(i+1)/2}}
    Y^{\floor{(i+1)/2}},
  \]
  which is co-nice. To establish the fast convergence, let
  $r\in\Real$. Choose $n_0\ge 2$ such that for $n\ge n_0$ and
  $m\in\{n-1,n+1\}$ it holds that $\big|f(m,n)-\Fp(m,n) \big|\le
  n^{-(r+2)}$. Then
  \[
    \big|m\cdot f(m,n)-m\cdot\Fp(m,n) \big|\le (n+1) \big|f(m,n)-\Fp(m,n) \big|\le
    \frac{n+1}{n^{r+2}}\le\frac{1}{n^r}.
  \]
  The second assertion in \ref{it:pFC2} follows from the first.

  Assertion~\ref{it:pFC3} follows from the observation that linear
  combinations of nice polynomials are nice.

  Assertion~\ref{it:pFC4} follows directly from \ref{it:pFC1} and \ref{it:pFC3}.
\end{proof}

\begin{remark}
  The proof of Lemma~\ref{lem:preFC}\ref{it:pFC2} is the only place
  where we need ``fast'' convergence, with a convergence rate bound by
  an inverse polynomial function in $n$, instead of just ordinary convergence.
\end{remark}

\begin{restatable}{lemma}{lemrelu}
  \label{lem:relu}
  The $\relu$ function preserves fast convergence.
  \end{restatable}
  The above lemma is relatively straightforward since $\relu$ is
  simply the identity in the positive half and a constant $0$ in the
  non-positive half.  

\begin{lemma}\label{lemma:ga_sigmoid}
The logistic function preserves fast convergence.
\end{lemma}

\begin{proof}
    Let $f\in\FC(\Fp)$ for some nice polynomial $\Fp$. (The proof for co-nice polynomials is similar.)
    We shall prove that $g:=\sig(f)\in \FC$. Note that
    $g(m,n)\coloneqq \sig(f(m,n))=\frac{1}{1+e^{-f(m,n)}}$.

    Let $a$ be the leading coefficient of $\Fp$.

    \smallskip\noindent
    \textit{Case 1:}
      $\Fp$ is not constant and $\sign(a)>0$.\\
      We shall prove that $g\in\FC(1)$.
      
      Let $r\in\Real$. Then for suffficiently
      large $n$ and $m\in\{n-1,n+1\}$ we have $f(m,n)\ge \frac{\Fp(m,n)}{2}\ge
      \frac{a}{4}n$ (since $f\in\FC(\Fp)$ and by
      Lemma~\ref{lem:nice-sign}) and 
      thus
      \begin{align*}
        \big|g(m,n)-1\big|
        &=\left|\frac{1}{1+e^{-f(m,n)}}-1\right|
        =\frac{e^{-f(m,n)}}{1+e^{-f(m,n)}}\\
        &\le e^{-f(m,n)}\le e^{-\frac{|a|}{4} n}\le\frac{1}{n^r}.
      \end{align*}
      \textit{Case 2:}
      $\Fp$ is not constant and $\sign(a)<0$.\\
      In this case, we can prove, similarly to Case~1, that
      $g\in\FC(0)$.

      \smallskip\noindent \textit{Case 3:}
      $\Fp$ is constant.\\
      Then $\Fp(m,n)=a$ for all $m,n$. We shall prove that
      $g\in\FC(\sig(a))$.
      For $n\ge 0$, we let
      \[
        \epsilon_{n}\coloneqq \max\big\{\abs{f(n-1,n)-a},
        \abs{f(n+1,n)-a}\big\}.
      \]
      Since $f\in\FC(\Fp)=\FC(a)$, for every $r\in\Real$ there is an $n_0$
      such that $\epsilon_n\le n^{-r}$ for all $n\ge n_0$.

      Let $n\ge n_0$ and $m\in\{n-1,n+1\}$. 
      We have $a-\epsilon_n\le f(m,n)\le
      a+\epsilon_n$ and thus 
      $
        \sig(a-\epsilon_n)\le g(m,n)\le\sig(a+\epsilon_n)
      $

      Since $\frac{d\,\sig}{d x}=\sig(x)(1-\sig(x))\le 1
      $ it follows that
      \begin{align*}
        &\big|g(m,n)-\sig(a)\big|\\
        &\le\max\Big\{\big|\sig(a-\epsilon_n)-\sig(a)\big|,
                                  \big|\sig(a+\epsilon_n)-\sig(a)\big|\big\}\\
        &\le\epsilon_n\le n^{-r}.
      \end{align*}
      This proves that $g\in\FC\big(\sig(a)\big)$.
\end{proof}

 In view of Lemma~\ref{lem:relu}, the following theorem immediately
 implies  Theorem~\ref{theo:unisep}.
 
\begin{theorem}\label{theo:unisep-gen}
  The query $\CQ_1$ is expressible by a 2-GNN with \SUM-aggregation,
  but not by a 1-GNN with \SUM-aggregation and with arbitrary
  activation functions that preserve fast convergence.
\end{theorem}

\begin{proof}
  Suppose for contradiction that
  \[
    \FN=(\FL^{(1)},\ldots,\FL^{(d)})
  \]
  is a 1-GNN
  that computes $\CQ_1$. By adding a bias of $-1/2$ to the output of
  the last layer, we may actually assume that for all graphs $G$ and
  vertices $v\in V(G)$ it holds that
  \begin{equation}
    \label{eq:30}
    S_{\FN}(G)(v)
    \begin{cases}
      \ge\frac{1}{4}&\text{if }v\in \CQ_1(G),\\
      \le-\frac{1}{4}&\text{if }v\not\in\CQ_1(G).
    \end{cases}
  \end{equation}
  By Lemma~\ref{lem:msg}, we may assume that the message function on
  each layer is the identity function.  Suppose that
  \[
    \FL^{(i)}=(\id_{p^{(i)}},\SUM,\comb^{(i)}),
  \]
  where $\id_{p^{(i)}}$
  is the identity on $\Real^{p^{(i)}}$ and
  $\comb^{(i)}:\Real^{2p^{(i)}}\to\Real^{q^{(i)}}$ is a function
  computed by a feed-forward neural network. Here
  $p^{(i)},q^{(i)}\in\Real$ are the input and output dimensions of the
  layer. We have $p^{(0)}=0$, because the input graph is unlabelled,
  $q^{(i)}=p^{(i+1)}$ for $0\le i<d$, and $q^{(d)}=1$. 

  As input graphs, we consider complete bipartite graphs
  $K_{m,n}$. We assume that
  $V(K_{m,n})=U\cup V$ for disjoint sets $U,V$ of sizes $|U|=m,|V|=n$ and
  $E(K_{m,n})=\{uv\mid u\in U,v\in V\}$.
  Note that
  \begin{equation}
    \label{eq:31}
    \CQ_1(K_{m,n})=
    \begin{cases}
      U&\text{if }m<n,\\
      \emptyset&\text{if }m=n,\\
      V&\text{if }m>n.
    \end{cases}
  \end{equation}
  For $m,n\in\Nat$, let $\Cf^{(0)}_{m,n}\in\CS_0(K_{m,n})$ be the $0$-dimensional signal
  on $K_{m,n}$ that maps each vertex to the empty tuple $()$, and for
  $1\le i\le d$, let $\Cf^{(i)}_{m,n}\coloneqq
  S_{\FL^{(i)}}(K_{m,n},\Cf^{(i-1)}_{m,n})$. Then  $\Cf^{(i)}_{m,n}\in\CS_{q^{(i)}}(K_{m,n})$
  is the signal we obtain by applying the first $i$ layers of $\FN$ to
  $K_{m,n}$. Since the transformations computed by GNNs are
  equivariant, for all $u,u'\in U$ it holds that
  $\Cf^{(i)}_{m,n}(u)=\Cf^{(i)}_{m,n}(u')$, and for all
  $v,v'\in V$ it holds that
  $\Cf^{(i)}_{m,n}(v)=\Cf^{(i)}_{m,n}(v')$.

  For $i\in[d],j\in[q^{(i)}]$, let $f^{(i)}_j:\Nat^2\to\Real$
  be the function defined by
  \[
    \Cf^{(i)}_{m,n}(u)=\big(f^{(i)}_1(m,n),\ldots,
    f^{(i)}_{q^{(i)}}(m,n)\big)
  \]
  for all $u\in U$.  That is, $f^{(i)}_j(m,n)$ is the $j^{th}$ entry of
  the state of a vertex $u\in U$ after applying the first $i$ layers of
  $\FN$ to a graph $K_{m,n}$. Similarly, we let
  $g^{(i)}_j:\Nat^2\to\Real$ be the function defined by
  \[
    \Cf^{(i)}_{m,n}(v)=\big(g^{(i)}_1(m,n),\ldots,
    g^{(i)}_{q^{(i)}}(m,n)\big)
  \]
  for all $v\in V$.

  \begin{claim}
    For all $i\in[d]$, $j\in[q^{(i)}]$ we have
    \[
      f^{(i)}_j\in\FC,\quad g^{(i)}_j\in\FCco.
    \]

    \proof
    The proof is by induction on $i$. 

    For the base step, we observe that $f^{(1)}$ must be constant,
    because $\Cf^{(0)}$ maps all vertices to the empty tuple, the
    message function is the identify, the sum of empty tuples is the
    empty tuples, so the combination function receives the empty tuple
    as input (at all vertices).


    For the inductive step, suppose that $i>1$ and that
    $f^{(i-1)}_j\in\FC,g^{(i-1)}_j\in\FCco$. Let $p\coloneqq
    p^{(i)}=q^{(i-1)}$, and let $q\coloneqq q^{(i)}$. It will be convenient to
    think of $\comb^{(i)}:\Real^{2p}\to\Real^q$ as a tuple
    $(c_1,\ldots,c_q)$, where $c_j:\Real^{2p}\to\Real^q$.
    Since
    $\comb^{(i)}$ is computable by an FNN with activations that
    preserv fast convergence,
    each $c_j$ preserves fast convergence. For all $u\in U$, we have
    \[
      \Cf_{m,n}^{(i)}(u)=\comb^{(i)}\left(\Cf^{(i-1)}(u),\sum_{v\in
          V}\Cf^{(i-1)}_{m,n}(v)\right)
    \]
    and thus, for $j\in[q]$,
    \begin{multline*}
      f^{(i)}_j(m,n)=c_j\Big(f^{(i-1)}_1(m,n),\ldots,
      f^{(i-1)}_p(m,n),\\
      n\cdot g^{(i-1)}_1(m,n),\ldots,
      n\cdot g^{(i-1)}_p(m,n)\Big).
    \end{multline*}
    By the induction hypothesis, we have $f^{(i-1)}_k\in\FC$ and
    $g^{(i-1)}_k\in\FCco$. The latter implies, by
    Lemma~\ref{lem:preFC}\ref{it:pFC2}, that the function
    $(m,n)\mapsto n\cdot g^{(i-1)}_k(m,n)$ is in $\FC$. Thus $f^{(i)}_j\in\FC$.

    The argument for $g^{(i)}_j$ is similar. This completes the proof
    of the claim.
    \uend    
  \end{claim}

  The function $f_1^{(d)}$ 
  computes the output of $\FN$ at the
  vertices in $U$: for all $u\in U$ we have
  $
    f^{(d)}_1(m,n)=\Cf^{(d)}_{m,n}(u)=S_{\FN}(K_{m,n})(u).
  $
  Thus by \eqref{eq:30} and \eqref{eq:31},
  \[
    f^{(d)}_1(m,n)
    \begin{cases}
      \ge\frac{1}{4}&\text{if }m<n,\\
      \le-\frac{1}{4}&\text{if }m\ge n.
    \end{cases}
  \]
  Thus for all $n\in\PNat$, $f^{(d)}(n-1,n)\ge\frac{1}{4}$ and
  $f^{(d)}(n+1,n)\le-\frac{1}{4}$. Since $f^{(d)}_1(m,n)\in\FC$, this contradicts Lemma~\ref{lem:nice-sign}.
\end{proof}

Interestingly, there are also somewhat natural functions that, when
used as activations, allow 1-GNNs to express the query $\CQ_1$.

\begin{example}\label{exa:act}
  \begin{enumerate}
  \item
  The query $\CQ_1$ can be expressed by a 1-GNN with
    $\SUM$-aggregation and the square root function as activation
    function.
  \item The main results of \cite{Grohe23} hold for a class of
    \emph{piecewise linear approximable} activations. The
    functions $f,g:\Real\to\Real$ defined by $f(x)=\min\{1,1/|x|\}$ and 
    $g(x)=\min\{1,1/\sqrt{|x|}\}$ (with $f(0)=g(0)=1$) are piecewise
    linear approximable. Yet it can be shown that $\CQ_1$ can be expressed by a 1-GNN with
    $\SUM$-aggregation that uses $f$ and $g$ as activations.   
  \end{enumerate}
  We leave the proofs of these assertions to the reader.
\end{example}

\subsection{The MEAN-MAX theorem}

Maybe surprisingly, the use of \SUM-aggregation is crucial in
Theorem~\ref{theo:unisep}. The corresponding result for \MEAN\ or
\MAX\ aggregation does not hold. To the contrary, we have the
following. 

\begin{restatable}{theorem}{thmmeanmax}\label{thm:mean-max}
  \begin{enumerate}  
  \item Every query computable by a 2-GNN with \MAX-aggregation is
    computable by a 1-GNN with \MAX-aggregation.
  \item Every query computable by a 2-GNN with \MEAN-aggregation is
    computable by a 1-GNN with \MEAN-aggregation.
  \end{enumerate}
\end{restatable}

To explain the proof, let us first consider \MAX-aggregation. Consider
a 2-GNN $\FN$ with \MAX-aggregation that computes some query $\CQ$ on
$\ell$-labelled graphs $(G,\Cf)\in\CGS_\ell^{\bool}$,
for some fixed $\ell$. The initial signal $\Cf$ only takes values
in the finite set $\{0,1\}^\ell$. It is not hard to prove, by
induction on the number of layers, that in a
2-GNN with \MAX-aggregation, the range of the signal remains finite
through all layers of the computation. Now the trick is to use a one-hot encoding of the
possible values. Suppose that after $i$-steps of the computation, the signal
$\Cf^{(i)}$ maps all vertices to vectors in some finite set
$S^{(i)}=\{\vec s_1,\ldots,\vec s_m\}\subseteq\Real^{q}$. Importantly, this set $S^{(i)}$
is independent of the input graph $(G,\Cf)\in\CGS_\ell^{\bool}$, it
only depends on $\ell$ and the GNN. In a one-hot encoding of the set
$S^{(i)}$ we represent the element $\vec s_j$ by the $j$-th unit vector
$\vec e_j\in\Real^m$. We
can construct a 1-GNN $\FN'$ that simulates the computation of
$\FN$, but represents all states $\Cf^{(i)}(u)$ by their one-hot
encoding. We only need a 1-GNN with identity messages for this,
because aggregating, that is, taking the coordinatewise maximum of,  
the one-hot encoded states of the neighbours
gives a node the full set of states appearing at the neighbours. Since
\MAX-aggregation is not sensitive to multiplicities, this is
sufficient to reconstruct the messages of the original 2-GNN $\FN$ and
simulate the necessary computations in the combination function.

If we try to prove the assertion for \MEAN-aggregation similarly, we
already fail at the first step. The range of the signals computed by a
\MEAN-GNN on inputs $(G,\Cf)\in\CGS_\ell^{\bool}$ is infinite in
general. The
trick to resolve this is to use finite-valued approximations and then
use one-hot encodings of these. The reason that this scheme works is that the functions computed by FNNs are Lipschitz continuous, and that
$\MEAN$ aggregation (as opposed to $\SUM$ aggregation) keeps the approximation error bounded. 

\section{Conclusions}

We study the expressivity of graph neural networks with 1-sided and
2-sided message passing, and closely related to this, the expressivity of modal and guarded first-order logic with
counting. The picture we see is surprisingly complicated: on the
logical side, the modal and guarded fragments have the same
expressivity. This implies that the 1-GNNs and 2-GNNs have the same
expressivity in non-uniform settings. This contrasts with the uniform
setting, where we can separate 1-GNNs from 2-GNNs, but only if they
use SUM aggregation.

The proofs of these results introduce novel techniques. Our proof that
$\MFOC\equiv\GFOC$ is based on a hashing trick; we are not aware of
similar arguments used elsewhere in a logical context (similar
techniques are being used in complexity theory). Furthermore, there
are not many inexpressibility results for graph neural networks that
are not based on the Weisfeiler-Leman algorithm. A notable exception
is \cite{RosenbluthTG23}, and we built on techniques developed
there. A specific idea that is new here is our notion of fast
convergence, which allows us to extend our result to activation
functions other than $\relu$.

Some questions remain open. In the uniform setting, the query that we use to separate 2-GNNs
from 1-GNNs with \SUM-aggregation turns out to be expressible by 1-GNNs that use
both \SUM\ and \MEAN. Is there also a query that is expressible by
2-GNNs, but not by 1-GNNs with \SUM\ and \MEAN, or even \SUM, \MEAN,
and \MAX\ aggregation?

In the non-uniform setting, we have not considered families of GNNs of
polynomial size and unbounded depth. Similarly to polynomial-size families of
Boolean circuits, such families can be very powerful, and we currently
have no techniques for establishing lower bounds (that is, inexpressivity)
except for arguments based on Weisfeiler-Leman, which cannot separate
1-GNNs from 2-GNNs. Related to this is also the question if recurrent
2-GNNs are more expressive than recurrent 1-GNNs.

\printbibliography

\pagebreak
\appendix
\renewcommand{\theequation}{\Alph{section}.\Alph{equation}}

\section{Details omitted from Section~\ref{sec:counting-logics}}

We start by proving \cref{prop:MCvsGC}. For the reader's convenience,
we restate the proposition, and we will do the same with all results
we prove in the appendix.

\propmcvsgc*

\begin{proof}
  We only need to translate each $\GC$-formula
  \[
    \phi(x)=\exists^{\ge n}x'. \big(E(x,x')\wedge\psi(x,x')\big),
  \]
  where (by induction) we assume that $\psi(x,x')$ is modal, to an
  $\MC$-formula.

  We note that $\psi(x,x')$ is a Boolean combination of formulas
  $x=x'$, $E(x,x')$ as well as $\MC$-formulas 
  with only one free variable. We may further assume that this Boolean
  combination is in disjunctive normal form, that is,
  $\psi(x,x')=\bigvee_{i=1}^k\gamma_i(x,x')$, where each $\gamma_i$ is
  a conjunction of atomic formulas $x=x',E(x,x')$, their negations
  $\neg x=x',\neg E(x,x')$, and $\MC$-formulas 
  with only one free variable. In fact, we may even assume that the
  $\gamma_i$ are mutually exclusive, that is, for $i\neq i'$ the
  formula $\gamma_i\wedge \gamma_{i'}$ is
  unsatisfiable. To achieve this, let us assume that
  $\gamma_{i}=(\lambda_1\wedge\ldots\wedge\lambda_m)$. Then we replace
  $\gamma_{i}\vee\gamma_{i'}$ by
  $\gamma_i\vee\bigvee_{j=1}^m(\gamma_{i'}\wedge\neg\lambda_j)$. We
  can easily lift this construction to the whole disjunction
  $\bigvee_{i=1}^k\gamma_i$.

  Since we are only considering
  simple undirected graphs, we do not need to consider atoms $E(x',x)$
  or $E(x,x)$. Of course we may assume that we never
  have an atom and its negation appearing together in the same
  conjunction $\gamma_i$. Again since we are in simple graphs, we may further assume that $x=x'$ and $E(x,x')$ never
  appear together in a $\gamma_i$.
Finally, since we are taking the conjunction of $\psi(x,x')$
  with $E(x,x')$, we may assume that $\neg E(x,x')$ and $x=x'$ never
  appear in a $\gamma_i$. Moreover, we can omit $E(x,x')$ and $\neg
  x=x'$ from all $\gamma_i$. All this implies that we may actually
  assume that $\gamma_i=\chi_i(x)\wedge\chi_i'(x')$ for $\MC$-formulas
  $\chi_i,\chi_i'$. Hence
  \[
    \phi(x)=\exists^{\ge
      n}x'. \left(E(x,x')\wedge\bigvee_{i=1}^k\big(\chi_i(x)\wedge\chi_i'(x')\big)\right).
  \]
  Then
  \begin{align*}
    \phi(x)&\equiv \exists^{\ge
             n}x'. \bigvee_{i=1}^k\big(E(x,x')\wedge\chi_i(x)\wedge\chi_i'(x')\big)\\
    &\equiv\bigvee_{\substack{n_1,\ldots,n_k\in\set{n+1}\\n_1+\ldots+n_k=n}}\bigwedge_{ji=1}^k \exists^{\ge
    n_j}x'.\big(E(x,x')\wedge\chi_i(x)\wedge\chi_i'(x')\big)\\
    &\equiv\bigvee_{\substack{n_1,\ldots,n_k\in\set{n+1}\\n_1+\ldots+n_k=n}}\bigwedge_{i=1}^k\Big(\chi_i(x)\wedge\exists^{\ge
             n_j}x'.\big(E(x,x')\wedge\chi_i'(x')\big)\Big).
  \end{align*}
  The last formula is modal.
\end{proof}

\lemsimplenf*

\begin{proof}
  To establish property (i) of simple formulas, we observe that
  \[
    \#(x_1,\ldots,x_k,y_1<\theta_1,\ldots,y_\ell<\theta_\ell).\psi
  \]
  is equivalent to
  \[
    \#(x_1,\ldots,x_k,y_1<\ord^{d_1},\ldots,y_\ell<\ord^{d_\ell}).\left(\bigwedge_{j=1}^\ell
      y_i<\theta_i\wedge \psi\right)
  \]
  if we can guarantee that $\theta_i$ takes values $< n^{d_i}$ in graphs
  of order $n$.

  Property (ii) is proved by an induction based on the following
  observation. Suppose that $\phi$ is a formula that contains a term
  $\theta$ such that no free variable of $\theta$ appears in the scope
  of any quantifier (i.e., counting operator) in $\phi$. To achieve
  this we can rename bound variables. Let $\phi'$ be the formula
  obtained from $\phi$ by replacing each occurrence of $\theta$ by a
  number variable $y$. Then $\phi$ is equivalent to
  $\exists y<\ord^d\big(y=\theta\wedge \phi')$ if we can guarantee
  that $\theta$ takes values $< n^d$ in graphs of order $n$.
\end{proof}

\lemhash*

\begin{proof}
  Suppose that $\CM=\{M_1,\ldots,M_n\}$ such that
  $M_1<M_2<\ldots<M_n$, and for $1\le i<j\le n$, let $D_{ij}\coloneqq
  M_j-M_i$. Note that for every $p$,
  \[
    M_i\equiv M_j\bmod p\iff p\bigmid D_{ij}.
  \]
  ($p\,|\,D_{ij}$ means ``$p$ divides $D_{ij}$''). Since
  $D_{ij}<2^m$, it has less than $m$ prime factors. Thus
  \[
    \Pr_{p\in P}\big(M_i\equiv M_j\bmod p)=\Pr_{p\in P}(p\,|\,D_{ij})<
    \frac{m}{|P|}.
  \]
  By the Union bound,
  \begin{align*}
    &\Pr_{p\in P}\big(\exists M,N\in\CM,M\neq N:\,M\equiv N\bmod p\big)\\
    =\,&\Pr_{p\in P}\big(\exists i,j\in[n],i<j:\,M_i\equiv M_j\bmod p)\\
    <\,&\frac{mn(n-1)}{2|P|}<\frac{1}{k}.
  \end{align*}
\end{proof}

\lemprimes*

\begin{proof}
  Let $\pi(n)$ be the number of primes in $[n]$.
  By the prime number theorem, we have
  \[
    \lim_{n\to\infty}\frac{\pi(n)}{n/\ln n}=1.
  \]
  Thus there is an $n_1$ such that for all $n\ge n_1$ we have
  $\pi(n)\ge \frac{3n}{4\ln n}$. Moreover, there is an $n_2\in\Nat$
  such that for all $n\ge n_2$ we have $2\ln (2\ln n)\le\ln n$. Then
  for $n\ge\max\{n_1,n_2\}$ we have 
  \[
    \pi(2n\ln n)\ge \frac{6n\ln n}{4\ln n+4\ln(2\ln n)}\le \frac{6n\ln
      n}{6\ln n}=n.
  \]
\end{proof}

\section{Details omitted from Section~\ref{sec:gnn}}

\lemmsg*

\begin{proof}
  The trick is to integrate the message function of layer $i$ into the
  combination function of layer $i-1$. This is why we need an additional $0$th
  layer. that does nothing but compute the message function of the
  first layer.
  
  Suppose that for $i\in[d]$ we have
  $\FL^{(i)}=(\msg^{(i)},\agg^{(i)},\comb^{(i)})$, where
  $\msg^{(i)}:\Real^{p^{(i)}}\to\Real^{r^{(i)}}$ and
  $\comb^{(i)}:\Real^{p^{(i)}+r^{(i)}}\to\Real^{q^{(i)}}$.

  We let $\tilde p^{(0)}\coloneqq p^{(1)}$ and
  $\tilde p^{(i)}\coloneqq p^{(i)}+r^{(i)}$ for $i\in[d]$. Moreover,
  for $i\in\set d$ we let $\tilde q^{(i)}\coloneqq \tilde p^{(i+1)}$,
  and we let $\tilde q^{(d)}\coloneqq q^{(d)}$.  For each $p\in\Nat$
  we let $\logic{id}_p$ be the identity function on $\Real^p$. For
  $i\in\set{d+1}$, we let
  $\tilde{\FL}^{(i)}=(\logic{id}_{\tilde p^{(i)}},\tilde{\agg}^{(i)},
  \tilde{\comb}^{(i)})$, where $\tilde{\agg}^{(0)}$ is arbitrary (say,
  \SUM), $\tilde{\agg}^{(i)}\coloneqq\agg^{(i)}$ for $i\in[d]$, and
  $\tilde{\comb}^{(i)}:\Real^{2\tilde p^{(i)}}\to\Real^{\tilde
    q^{(i)}}$ is defined as follows:
  \begin{itemize}
  \item for $i=0$ and $\vec x,\vec z\in\Real^{p^{(1)}}$ we let
      \[
        \tilde{\comb}^{(0)}(\vec x,\vec z)\coloneqq\big(\vec
        x,\msg^{(1)}(\vec z)\big);
      \]
    \item for $1\le i\le d-1$  and $\vec x,\vec
      x'\in\Real^{p^{(i)}},\vec z,\vec z'\in\Real^{r^{(i)}}$ we let
     \begin{multline*}
        \hspace{10mm}\tilde{\comb}^{(i)}(\vec x\vec z,\vec x'\vec
        z')\coloneqq\\
        \Big(\comb^{(i)}(\vec x,\vec z'),\msg^{(i+1)}\big(\comb^{(i)}(\vec x,\vec z')\big)\Big);
      \end{multline*}
    \item
        for $i=d$  and $\vec x,\vec
      x'\in\Real^{p^{(d)}},\vec z,\vec z'\in\Real^{r^{(d)}}$ we let
     \[
        \tilde{\comb}^{(d)}(\vec x\vec z,\vec x'\vec
        z')\coloneqq
        \comb^{(d)}(\vec x,\vec z').
      \]
  \end{itemize}
  It is straightforward to verify that $T_{\FN}=T_{\tilde\FN}$.
\end{proof}

\proplinearmsg*

\begin{proof}
  It suffices to prove this for a single layer. So let
  $\FL=(\msg,\MEAN,\comb)$ be a 2-GNN layer of input dimension $p$,
  output dimension $q$ where $\msg:\vec x\mapsto A\vec x+\vec b$  
  where
  $A\in\Real^{r\times 2p},\vec b\in\Real^r$. For all vectors
  $\vec x,\vec x_1,\ldots,\vec x_n\in\Real^p$ we have
  \[
    \agg\Big(\biglmulti A(\vec x,\vec x_i)+\vec b\bigmid
    i\in[n]\bigrmulti\Big)=A\left(\vec x,\agg\big(\lmulti\vec
      x_i\mid i\in[n]\rmulti\right)+\vec b.
  \]
  We define $\tilde{\comb}:\Real^{2p}\to\Real^q$ by
  \begin{equation}
    \label{eq:9}
    \tilde{\comb}(\vec x,\vec x')=\comb\left(\vec x, A\left(\vec 
        x,\vec x'\right)+\vec b\right). 
  \end{equation}
  Note that this function is computable by an FNN (assuming $\comb$
  is). We let
  $\tilde{\FL}\coloneqq(\logic{id}_p,\MEAN,\tilde{\comb})$. Then
  $T_{\FL}=T_{\tilde\FL}$.
\end{proof}

Observe that if the GNN layer $\FL$ in the previous proof had $\SUM$
instead of $\MEAN$ as aggregation, we could still write down the
function corresponding to \eqref{eq:9}, but in general it would not be
comoputable by an FNN.

\section{Details omitted from Section~\ref{sec:uniform}}

\subsection{SUM}





\lempreFC*

\begin{proof}
  All polynomials of degree $0$ (constants) are nice, which proves
  \ref{it:pFC1}.

  To prove~\ref{it:pFC2}, suppose that $f\in\FC(\Fp)$ for the polynomial $\Fp$
  in \eqref{eq:1}. Note that
  \[
    X\cdot\Fp(X,Y)=\sum_{i=0}^k a_i X^{\floor{i/2}+1}
    Y^{\ceil{i/2}}=\sum_{i=0}^k a_i X^{\ceil{(i+1)/2}}
    Y^{\floor{(i+1)/2}},
  \]
  which is co-nice. To establish the fast convergence, let
  $r\in\Real$. Choose $n_0\ge 2$ such that for $n\ge n_0$ and
  $m\in\{n-1,n+1\}$ it holds that $\big|f(m,n)-\Fp(m,n) \big|\le
  n^{-(r+2)}$. Then
  \[
    \big|m\cdot f(m,n)-m\cdot\Fp(m,n) \big|\le (n+1) \big|f(m,n)-\Fp(m,n) \big|\le
    \frac{n+1}{n^{r+2}}\le\frac{1}{n^r}.
  \]
  The second assertion in \ref{it:pFC2} follows from the first.

  Assertion~\ref{it:pFC3} follows from the observation that linear
  combinations of nice polynomials are nice.

  Assertion~\ref{it:pFC4} follows directly from \ref{it:pFC1} and \ref{it:pFC3}.
\end{proof}

\lemrelu*

\begin{proof}
  Suppose that $f\in\FC(\Fp)$ for the polynomial $\Fp$ in
  \eqref{eq:1}. We need to show that $\relu(f)\in \FC$. Then the
  assertion for co-nice polynomials follows by flipping the
  arguments. Let $a$ be the leading coefficient of $\Fp$. If $a=0$,
  then $\Fp=0$ is the zero polynomial, and $\relu(f)\in\FC(\Fp)$
  follows from the observation that $|\relu(f(m,n))|\le|f(m,n)|$ for
  all $m,n$. Suppose that $a\neq 0$. By Lemma~\ref{lem:nice-sign},
  there is an $n_0$ such that for all $n\ge n_0$ and $m\in\{n-1,n+1\}$
  we have
  \[
    \Fp(m,n)
    \begin{cases}
      \ge a&\text{if }a>0,\\
      \le a&\text{if }a<0.
    \end{cases}
  \]
  By fast convergence, we may further assume that $|f(m,n)-\Fp(m,n)|\le\frac{a}{2}$.
  Thus if $a>0$, we have $\relu(f(m,n))=f(m,n)$ and thus
  $\relu(f)\in\FC(\Fp)$. If $a<0$ we have $\relu(f(m,n))=0$ and thus $\relu(f)\in\FC(0)$.
\end{proof}

\subsection{MEAN and MAX}

The rest of the appendix is devoted to a proof of the following theorem.

\thmmeanmax*
We introduce additional notations that we use in the proof.
For $n\in\PNat$, denote by $\vec1_{n,i}$ the vector of length $n$ with $1$ in the i$^{th}$ position and $0$ elsewhere. Let $\CX\subseteq\Real^{p}$ be a finite set and assume an enumeration of it $\CX=\vec x_1,\ldots,\vec x_n$. For all $i\in[n]$  denote by $\vec 1_{\CX}(\vec x_i)\coloneqq \vec1_{n,i}$ the one-hot representation of $\vec x_i$. Denote by $\vec 1_{\CX}^{-1}$ the inverse function of $\vec 1_{\CX}$, that is, $\vec 1_{\CX}^{-1}(1_{n,i})\coloneqq \vec x_i$. For a multiset $M\in\multiset{\CX}{*}$ we define $\vec 1_{\CX}(M)\coloneqq\biglmulti \vec 1_\CX(\vec x) \bigmid \vec x\in M\bigrmulti$ the corresponding multiset of one-hot representations.
For every two vectors $\vec u=(u_1,\ldots,u_k),\vec v=(v_1,\ldots,v_k)\in\Real^k$ we define $\|\vec u-\vec v\|\coloneqq \|\vec u-\vec v\|_\infty=\max_{i\in[k]}|u_i-v_i|$. Moreover, we let $\mean(\vec u)\coloneqq\frac{1}{k}\sum u_i$ and $\max(\vec u)=\max_{i\in[k]}u_i$.
We define an operator $U$ that removes the duplicates in a multiset, that is, for a multiset $M$ we have $U(M)\coloneqq\{x : x\in M\}$.

Let $\FN=(\FL^{(1)},\ldots,\FL^{(d)})$ be a $d$-layer i-GNN and let $p^{(i)},q^{(i)}$ be the input and output dimensions of $\FL^{(i)}$. For every graph and signal $(G,\Cf)\in\CGS_{p^{(1)}}$ and $i\in[d]$, we define 
\[T_\FN^{(i)}(G,\Cf)\coloneqq T_{\FL^{(i)}}\circ T_{\FL^{(i-1)}}\circ\ldots\circ T_{\FL^{(1)}}\]
the result of operating the first $i$ layers of $\FN$ on $(G,\Cf)$. We define the $0$-stage signal ${S_{\FN}^{(0)}(G,\Cf)\coloneqq \Cf}$, and for $i\in[d]$ we define the $i^{th}$-stage signal to be $S_{\FN}^{(i)}(G,\Cf)$ such that $$T_\FN^{(i)}(G,\Cf)=(G,S_{\FN}^{(i)}(G,\Cf))$$
When $(G,\Cf)$ are clear from the context, for a vertex $v\in V(G)$ we define $v^{(i)}_\FN\coloneqq S_{\FN}^{(i)}(G,\Cf)(v)$ the signal of $v$ after operating the first $i$ layers of $\FN$.
Also, when referring to a final signal value $S_{\FN}(G,\Cf)(v)$ we may omit $(G,\Cf)$ and write $S_{\FN}(v)$.
For every $(G,\Cf)\in\CGS_{p^{(1)}}$, every $v\in V(G)$, and every $i\in[d]$, we define 
$$N^{(i)}_{\FN}(v)\coloneqq\multi{w^{(i)}_\FN : w\in N_G(v)}$$ 
the multiset of signal-values of the neighbours of $v$ after the operation of the first $i$ layers of $\FN$.

\begin{proof}[Proof of Theorem~\ref{thm:mean-max}(1) (MAX-aggregation)]\mbox{}\\
Let $$\FN=(\FL^{(1)},\ldots,\FL^{(d)})$$ be a $d$-layer 2-GNN where $\FL^{(i)}=(\msg^{(i)},\max,\comb^{(i)})$, $\FL^{(i)}$ has input and output dimensions $p^{(i)},q^{(i)}$, and $\msg^{(i)}$ has output dimension $r^{(i)}$.
We define $D^{(0)}_\FN\coloneqq \{0,1\}^{p^{(1)}}$ the set of possible input boolean-signal values
and for $i\in[d]$ we define $$D^{(i)}_\FN\coloneqq \{S_{\FN}^{(i)}(G,\Cb)(v) : (G,\Cb)\in\CGS^\bool_{p^{(1)}}, v\in V(G)\}$$ the set of possible signal values after operating the first $i$ layers of $\FN$ when the initial input-signal is boolean. For $i\in[d+1[$ define $d^{(i)}_\FN\coloneqq \abs{D^{(i)}_\FN}\in(\Nat\cup\{\infty\})$ the size of $D^{(i)}_\FN$.
Assume $i\in[d[$, assume $D^{(i)}_\FN$ is finite, and assume an enumeration $D^{(i)}_\FN=x_1,\ldots,x_{d^{(i)}_\FN}$.

We define $\Fm_0:1_{D^{(i)}_\FN}(D^{(i)}_\FN)\times\{0,1\}^{d^{(i)}_\FN}\rightarrow D^{(i)}_\FN\times2^{D^{(i)}_\FN}$ such that 
$$\forall (s,t)\in 1_{D^{(i)}_\FN}(D^{(i)}_\FN)\times\{0,1\}^{d^{(i)}_\FN}$$
$$\Fm_0(s,t)\coloneqq (1^{-1}_{D^{(i)}_\FN}(s),\{x_i:t(i)=1\})$$
and note that for every $(G,\Cb)\in\CGS^\bool_{p^{(1)}}, v\in V(G)$
\begin{align*}
\Fm_0(1_{D^{(i)}_\FN}(v^{(i)}_{\FN}),\max(1_{D^{(i)}_\FN}(N^{(i)}_{\FN}(v))))
=(v^{(i)}_{\FN},U(N^{(i)}_{\FN}(v)))
\end{align*}
We define 
$\Fm_1:D^{(i)}_\FN\times2^{D^{(i)}_\FN}\rightarrow \Real^{r^{(i+1)}}$
such that
\begin{align*}
&\forall (G,\Cb)\in\CGS^\bool_{p^{(1)}}, v\in V(G)\quad \Fm_1(v^{(i)}_{\FN}, U(N^{(i)}_{\FN}(v)))\coloneqq
\\&\max(\msg^{(i+1)}(v^{(i)}_{\FN},x):x\in U(N^{(i)}_{\FN}(v))))
\end{align*}
Finally, we define
$\Fm_2:1_{D^{(i)}_\FN}(D^{(i)}_\FN)\times\{0,1\}^{d^{(i)}_\FN}\rightarrow\Real^{r^{(i+1)}}$
such that $$\Fm_2\coloneqq \Fm_1\circ \Fm_0$$
and have that for every $(G,\Cb)\in\CGS^\bool_{p^{(1)}}, v\in V(G)$
\begin{align*}
&\Fm_2(1_{D^{(i)}_\FN}(v^{(i)}_{\FN}),\max(1_{D^{(i)}_\FN}(N^{(i)}_{\FN}(v))))=
\\&\max(\msg^{(i+1)}(v^{(i)}_{\FN},x):x\in U(N^{(i)}_{\FN}(v)))
\\&=\max(\msg^{(i+1)}(v^{(i)}_{\FN},x):x\in N^{(i)}_{\FN}(v)))
\end{align*}
The last equality is key in determining that the domain of the $\comb$ function is finite although there are infinitely many possible multiset-inputs to the $\max$ aggregation.
By assumption that $D^{(i)}_\FN$ is finite, the domain of $\Fm_2$ is finite and there exists an FNN that implements $\Fm_2$. Denote that FNN by $\CF$, denote the function it implements by $f_{\CF}$, and note that $\comb^{(i+1)}$ is implemented by an FNN, then there exists an FNN that implements
$$comb'^{(i+1)}:\{0,1\}^{d^{(i)}_\FN}\times\{0,1\}^{d^{(i)}_\FN}\rightarrow D^{(i+1)}_\FN$$
such that for every $(G,\Cb)\in\CGS^\bool_{p^{(1)}}, v\in V(G)$
\begin{equation}\label{map3}
\begin{split}
&\comb'^{(i+1)}(1_{D^{(i)}_\FN}(v^{(i)}_{\FN}),\max(1_{D^{(i)}_\FN}(N^{(i)}_{\FN}(v))))\coloneqq
\\&\comb^{(i+1)}(v^{(i)}_{\FN},f_{\CF}(1_{D^{(i)}_\FN}(v^{(i)}_{\FN}),\max(1_{D^{(i)}_\FN}(N^{(i)}_{\FN}(v)))))=
\\&\comb^{(i+1)}(v^{(i)}_{\FN},\max(\msg(v^{(i)}_{\FN},x):x\in N^{(i)}_{\FN}(v)))=v^{(i+1)}_{\FN}
\end{split}
\end{equation}
Note that  (assuming $D^{(i)}_\FN$ is finite) the domain of $comb'^{(i+1)}$ is finite, hence $D^{(i+1)}_\FN$ is finite. Since $D^{(0)}_\FN$ is finite by definition, we have by induction that $D^{(i)}_\FN$ is finite $\forall i\in[d+1[$ and there exists $\comb'^{(i+1)}$ as described. For $i\in[d-1[$ we can modify the $\comb'^{(i+1)}$-implementing FNN into a $\comb''^{(i+1)}$-implementing FNN that outputs $1_{D^{(i+1)}_\FN}(v^{(i+1)}_{\FN})$ instead of $v^{(i+1)}_{\FN}$.
Let $\FL'^{(0)}$ be a 1-GNN layer (which clearly exists) such that for every ${(G,\Cb)\in\CGS^\bool_{p^{(1)}}, v\in V(G)}$ it holds that  $S_{\FL'^{(0)}}(G,\Cb)(v)= 1_{D^{(0)}_\FN}(\Cb(v))$, and define 
$\forall i\in[d-1]$ ${\FL'^{(i)}\coloneqq(\id_{d^{(i-1)}_\FN},\max,\comb''^{(i)})}$ and 
$${\FL'^{(d)}\coloneqq(\id_{d^{(d-1)}_\FN},\max,\comb'^{(d)})}$$
Finally, define
$$\FN'\coloneqq(\FL'^{(0)},\ldots,\FL'^{(d)})$$ then $T_{\FN'}=T_{\FN}$.
\end{proof}

\begin{proof}[Proof of Theorem~\ref{thm:mean-max}(2) (MEAN-aggregation)]\mbox{}\\
We define $\frac{1}{\PNat}\coloneqq\{\frac{1}{a}:a\in\PNat\}$.
For every $\delta\in\frac{1}{\PNat}$ and $n\in\Nat$, we define $$S_{n,\delta}\coloneqq\{(p_1,\ldots,p_n): \Sigma p_i=1,\;\forall i\in[n]\; \exists b\in\Nat : p_i=b\delta\}$$
the set of ${n+\frac{1}{\delta}-1 \choose  n-1}$ possible elements' proportions in a multiset over a set of size $n$, s.t. the proportion of each element is a multiple of $\delta$. 
Let $\FN=(\FL^{(1)},\ldots,\FL^{(d)})$ be a $d$-layer 2-GNN where $$\FL^{(i)}=(\msg^{(i)},\mean,\comb^{(i)})$$ 
and $\FL^{(i)}$ has input and output dimensions $p^{(i)},q^{(i)}$.
Let $\CX\in\multiset{\Real^{p^{(i)}}}{*}$ be a finite multiset, and assume an enumeration $\CX=x_1,\ldots,x_n$, then for $y\in\Real^{p^{(i)}}$ we denote by $$\Msg^{(i)}_\CX(y)\coloneqq (\msg^{(i)}(y,x_1),\ldots,\msg^{(i)}(y,x_{n}))$$ the vector of messages between $y$ and every element in $\CX$.
We prove by induction on $d$ that for every $\epsilon>0$ there exists a 1-GNN $\FN'=\FL'^{(0)},\FL'^{(1)},\ldots,\FL'^{(d)}$ and a finite set $\CX^{(d)}$ such that 
$$\forall(G,\Cb)\in\CGS^\bool_{p^{(1)}}\; \forall v\in V(G)\; \exists x_v \in\CX^{(d)} : $$
$$\abs{S_\FN(v)-x_v}<\frac{\epsilon}{2}\;,\;\abs{S_{\FN'}(v)-x_v}<\frac{\epsilon}{2}$$
which implies $$\abs{S_\FN(v)-S_{\FN'}(v)}<\epsilon$$
The layers $\FL'^{(i)}, i\in[d]$ of the 1-GNN are designed to operate on one-hot representations, and the layer $\FL'^{(0)}$ translates the initial-signal range to its one-hot representation.
The $\CX^{(i)}$s are the "make-believe" finite domains by which we will design the layers of the 1-GNN.
\subsubsection*{Induction Basis}    
We prove for $d=1$. Let $\epsilon>0$. 
Define ${\CX\coloneqq\{0,1\}^{p^{(1)}}}$, $n\coloneqq \abs{\CX}$, and assume an enumeration $\CX=x_1,\ldots,x_n$. Note that  
for every $(G,\Cb)\in\CGS^\bool_{p^{(1)}}, v\in V(G)$
\begin{equation}\label{eq:note}
    \begin{split}
&\mean(\msg^{(1)}(v^{(0)}_\FN,x):x\in N^{(0)}_\FN(v))=
\\&\Msg^{(1)}_\CX(v^{(0)}_{\FN})\cdot\mean(1_{\CX}(N^{(0)}_\FN(v)))
    \end{split}
\end{equation}
That is, the mean of the messages is equal to the product of the vector of possible messages and the vector of corresponding proportions of the elements (of $\CX$) among the neighbors of $v$.
By $\CX$ being finite, there exists ${g_1:\Real\rightarrow\Real}$, $\lim_{x\rightarrow0}g_1(x)=0$ such that $\forall (G,\Cb)\in\CGS^\bool_{p^{(1)}}\;\forall v\in V(G)\;\forall \delta>0\;\forall u\in\Real^{r^{(i)}}$
\begin{align*}    
    &\abs{u-\mean(1_{\CX}(N^{(0)}_\FN(v)))}\leq \delta \Rightarrow
        \\&|\Msg^{(1)}_\CX(v^{(0)}_{\FN})\cdot\mean(1_{\CX}(N^{(0)}_\FN(v))) - 
    \\&\Msg^{(1)}_\CX(v^{(0)}_{\FN})\cdot u|<g_1(\delta)
\end{align*}
Hence, by $\comb^{(1)}$ being Lipschitz-Continuous, and remembering \Cref{eq:note}, there exists ${g_2:\Real\rightarrow\Real}$, $\lim_{x\rightarrow0}g_2(x)=0$ such that
$\forall (G,\Cb)\in\CGS^\bool_{p^{(1)}}\;\forall v\in V(G)\;\forall \delta>0\;\forall u\in\Real^{r^{(i)}}$
\begin{equation}\label{eq:eq_1}
    \begin{split}
    &\abs{u-\mean(1_{\CX}(N^{(0)}_\FN(v)))}\leq \delta \Rightarrow
    \\& |\comb^{(1)}(v^{(0)}_{\FN},\mean(\msg^{(1)}(v^{(0)}_\FN,x):x\in N^{(0)}_\FN(v)))- 
            \\&\comb^{(1)}(v^{(0)}_{\FN},\Msg^{(1)}_\CX(v^{(0)}_{\FN})\cdot u)|<g_2(\delta_1)
    \end{split}                                       
\end{equation}
Let $\delta_1\in\frac{1}{\PNat}$ such that $\forall \delta\leq\delta_1\; g_2(\delta)<\frac{\epsilon}{2}$.
Let $(G,\Cb)\in\CGS^\bool_{p^{(1)}},v\in V(G)$ 
and let $\vec{p}_v\in S_{n,\delta_1}$ be proportions - which must exist - such that 
$$\abs{\vec{p}_v-\mean(1_{\CX}(N^{(0)}_\FN(v)))}\leq \delta_1$$
, then, 
\begin{equation}\label{eq:eq_1_5}
    \begin{split}
    & |\comb^{(1)}(v^{(0)}_{\FN},\mean(\msg^{(1)}(v^{(0)}_\FN,x):x\in N^{(0)}_\FN(v)))- 
    \\&\comb^{(1)}(v^{(0)}_{\FN},\Msg^{(1)}_\CX(v^{(0)}_{\FN})\cdot\vec{p}_v)|<\frac{\epsilon}{2}
    \end{split}                                       
\end{equation}    
Note that the first expression in the LHS of \Cref{eq:eq_1_5} is $S_\FN(v)$.
We proceed to show a similar result for $S_{\FN'}$.
For $\delta_2\in\frac{1}{\PNat}$ define $\comb'^{(1)}:\{0,1\}^n\times\Real^n \rightarrow \Real^{q^{(1)}}$ such that
$$\forall (x,\vec{p})\in 1_{\CX}(\CX)\times S_{n,\delta_2}\;\; \comb'^{(1)}(x,\vec{p})\coloneqq$$
$$\comb^{(1)}(1^{-1}_{\CX}(x),\Msg^{(1)}_\CX(1^{-1}_{\CX}(x)) \cdot \vec{p})$$
While the domain on which $\comb'^{(1)}$ is applicable is infinite, the domain for which it is explicitly specified is finite, hence 
$\comb'^{(1)}$ can be implemented by an FNN.
By argumentation similar to the development of \Cref{eq:eq_1} we have that 
there exists $$g_3:\Real\rightarrow\Real, \lim_{x\rightarrow0}g_3(x)=0$$ such that $\forall (G,\Cb)\in\CGS^\bool_{p^{(1)}}\;\forall v\in V(G)\;\forall \delta_2>0\;\forall u\in\Real^{r^{(i)}}$
\begin{equation}\label{eq:eq_2}
    \begin{split}
    &\abs{u-\mean(1_{\CX}(N^{(0)}_\FN(v)))}\leq \delta_2 \Rightarrow
     \\&|\comb'^{(1)}(1_{\CX}(v^{(0)}_{\FN}),\mean(1_{\CX}(N^{(0)}_\FN(v))))- 
      \\&\comb'^{(1)}(1_{\CX}(v^{(0)}_{\FN}),u)|<g_3(\delta_2)
    \end{split}                                       
\end{equation}

Let $\delta_2\in\frac{1}{\PNat}$ such that $\forall \delta\leq\delta_2\; g_3(\delta)<\frac{\epsilon}{2}$.
Let $(G,\Cb)\in\CGS^\bool_{p^{(1)}},v\in V(G)$ 
and let $\vec{p}_v\in S_{n,\delta_2}$ be proportions - which must exist - such that 
$$\abs{\vec{p}_v-\mean(1_{\CX}(N^{(0)}_\FN(v)))}\leq \delta_2$$
, then, 
\begin{equation}\label{eq:eq_2_5}
    \begin{split}
     &|\comb'^{(1)}(1_{\CX}(v^{(0)}_{\FN}),\mean(1_{\CX}(N^{(0)}_\FN(v))))- 
      \\&\comb'^{(1)}(1_{\CX}(v^{(0)}_{\FN}),\vec{p}_v)|<\frac{\epsilon}{2}
    \end{split}                                       
\end{equation}
Define $\FL'^{(0)}$ to be a 1-GNN layer (which clearly exists) such that $S_{\FL'^{(0)}}(G,\Cb)(v)\coloneqq 1_{\CX}(\Cb(v))$.      
Define $\FL'^{(1)}\coloneqq(\id_n,\mean,\comb'^{(1)})$ and define 
$$\FN'=(\FL'^{(0)},\FL'^{(1)})$$
Then, note that the first expression in the LHS of \Cref{eq:eq_2_5} is $S_{\FN'}(v)$.
Define $\delta_3=min(\delta_1, \delta_2)$, then by \Cref{eq:eq_1_5} and \Cref{eq:eq_2_5}, for every $(G,\Cb)\in\CGS^\bool_{p^{(1)}},v\in V(G)$ there exists $\vec{p}_v\in S_{n,\delta_3}$ such that for
$$x_v\coloneqq \comb^{(1)}(v^{(0)}_{\FN}, \Msg^{(1)}_\CX(v^{(0)}_{\FN})\cdot\vec{p}_v)$$
it holds that
$$|S_\FN(v) - x_v|<\frac{\epsilon}{2},\;\; |S_{\FN'}(v) - x_v|<\frac{\epsilon}{2}$$    
Hence, define 
$$\CX^{(1)}\coloneqq \{\comb^{(1)}(y, \Msg^{(1)}_\CX(y)\cdot\vec{p}) : (y,\vec{p})\in \CX\times S_{n,\delta_3}\}$$
then $\CX^{(1)}$ satisfies the requirement.

\subsubsection*{Induction Step}
We assume correctness for $d=t$ and every $\epsilon'>0$ and prove for $d=t+1$. Define $\CX\coloneqq \CX^{(t)}, n\coloneqq \abs{\CX}$, and assume an enumeration $\CX=x_1,\ldots,x_n$.
Let $\FN''=(\FL''^{(0)},\ldots,\FL''^{(t)})$
be a 1-GNN - that exists by the induction assumption - such that $$\forall(G,\Cb)\in\CGS^\bool_{p^{(1)}}\; \forall v\in V(G)\;  
\abs{v^{(t)}_{\FN}-S_{\FN''}(v)}<\epsilon'$$
and denote by $v^\CX$ a value in $\CX$ - which exists by the induction assumption - 
such that $\abs{v^{(t)}_{\FN}-v^\CX}<\frac{\epsilon'}{2}, \abs{S_{\FN''}(v)-v^\CX}<\frac{\epsilon'}{2}$. Compared to the case of $d=1$, in the case of $d>1$ we need to account not only for differences (which can be made as small as we want) between the actual proportions and the proportions that $\FL'^{(t+1)}$ will be designed for, but also for $\abs{v^{(t)}_{\FN}-v^\CX}$ i.e. the difference between the actual vertices' values and the values in $\CX$ - the domain that $\FL'^{(t+1)}$ will be designed for. We handle the proportions difference first, using the already walked-through path used in the proof for $d=1$.
For $\delta_1\in\frac{1}{\PNat}$ define $\comb'^{(t+1)}:\Real^n\times\Real^n \rightarrow \Real^{q^{(t+1)}}$ such that 
$$\forall (x,\vec{p})\in 1_{\CX}(\CX)\times S_{n,\delta_1}\;\; \comb'^{(t+1)}(x,\vec{p})\coloneqq$$
$$\comb^{(t+1)}(1^{-1}_{\CX}(x),(\Msg^{(t+1)}_\CX(1^{-1}_{\CX}(x))\cdot \vec{p}))$$
As explained in the case of $d=1$, such $comb'^{(t+1)}$ can be implemented by an FNN. Also, by the same argumentation used in the case of $d=1$, there exists $\delta_1\in\frac{1}{\PNat}$ such that for every $(G,\Cb)\in\CGS^\bool_{p^{(1)}},v\in V(G)$ there exists $\vec{p}_v\in S_{n,\delta_1}$ such that for
$$x_v\coloneqq \comb^{(t+1)}(v^{\CX},\Msg^{(t+1)}_\CX(v^{\CX})\cdot\vec{p}_v)$$
it holds that
\begin{equation}\label{eq:eq_3}
  \abs{\comb^{(t+1)}(v^\CX,\mean(\msg^{(t+1)}(v^\CX,w^\CX):w\in N_v(G)))-x_v}<\frac{\epsilon}{4}  
\end{equation}    
and     
\begin{equation}\label{eq:eq_4}
  \abs{\comb'^{(t+1)}(1_\CX(v^\CX),\mean(1_\CX(w^\CX):w\in N_v(G)))-x_v}<\frac{\epsilon}{4}  
\end{equation}

Next, we turn to the differences between the actual vertices' values (after layer $t$) and the values in $\CX$ - the finite domain according to which $\FL'^{(t+1)}$ is defined. We start with bounding the gap for the 2-GNN. 
By $\msg^{(t+1)}$ being Lipschitz-Continuous, there exists ${g_1:\Real\rightarrow\Real}$, $\lim_{x\rightarrow0}g_1(x)=0$ such that
$\forall (G,\Cb)\in\CGS^\bool_{p^{(1)}}\;\forall v,w\in V(G)$
$$\abs{\msg^{(t+1)}(v^{(t)}_{\FN},w^{(t)}_{\FN})-\msg^{(t+1)}(v^\CX,w^\CX)}<g_1(\epsilon')$$
, and thus, such that $\forall (G,\Cb)\in\CGS^\bool_{p^{(1)}}\;\forall v\in V(G)$
$$|\mean(\msg^{(t+1)}(v^{(t)}_{\FN},w^{(t)}_{\FN}):w\in N_v(G))-$$
$$\mean(\msg^{(t+1)}(v^\CX,w^\CX):w\in N_v(G))|<g_1(\epsilon')$$
By $\comb^{(t+1)}$ being Lipschitz-Continuous, we have that there exists $g_2:\Real\rightarrow\Real, \lim_{x\rightarrow0}g_2(x)=0$
such that
$$\forall (G,\Cb)\in\CGS^\bool_{p^{(1)}}\;\forall v\in V(G)$$
$$|\comb^{(t+1)}(v^{(t)}_{\FN},\mean(
\msg^{(t+1)}(v^{(t)}_{\FN},w^{(t)}_{\FN}):w\in N_v(G)))-$$ 
$$\comb^{(t+1)}(v^\CX,\mean(\msg^{(t+1)}(v^\CX,w^\CX):w\in N_v(G)))|<g_2(\epsilon')$$        
Assuming $\epsilon'$ small enough such that $\forall \epsilon''< \epsilon'\;\;g_2(\epsilon'')<\frac{\epsilon}{4}$ we have
\begin{equation}\label{eq:eq_5}      
  \begin{split}
  &\forall (G,\Cb)\in\CGS^\bool_{p^{(1)}}\;\forall v\in V(G)
  \\&|\comb^{(t+1)}(v^{(t)}_{\FN},\mean(\msg^{(t+1)}(v^{(t)}_{\FN},w^{(t)}_{\FN}):w\in N_v(G)))-
  \\&\comb^{(t+1)}(v^\CX,\mean(\msg^{(t+1)}(v^\CX,w^\CX):w\in N_v(G)))|<\frac{\epsilon}{4}
  \end{split}      
\end{equation}
Next, we bound the gap for the 1-GNN. Define $h:\CX\rightarrow 1_{\CX}(\CX)$ such that $\forall x\in\CX\;\; h(x)=1_{\CX}(x)$, a mapping from the values-domain to its one-hot representation, which can be implemented by an FNN.
Define $\FN'\coloneqq(\FL'^{(0)},\ldots,\FL'^{(t)})$ such that $\forall 0\leq i\leq (t-1)\; \FL'^{(i)}\coloneqq\FL''^{(i)}$, and 
$\FL'^{(t)}$ is such that 
$$S_{\FL'^{(t)}}\coloneqq h\circ S_{\FL''^{(t)}}$$
By the induction assumption, for every $(G,\Cb)\in\CGS^\bool_{p^{(1)}},v\in V(G)$ 
$\abs{S_{\FN''}(v)-v^\CX}<\frac{\epsilon'}{2}$. 
Hence, by (every FNN-implementation of) $h$ being Lipschitz-Continuous, there exists ${g_3:\Real\rightarrow\Real}$, ${\lim_{x\rightarrow0}g_3(x)=0}$ such that for every $(G,\Cb)\in\CGS^\bool_{p^{(1)}},v\in V(G)$ 
$$\abs{S_{\FN'}(v)-1_\CX(v^\CX)}<g_3(\epsilon')$$
, and thus, such that for every $(G,\Cb)\in\CGS^\bool_{p^{(1)}},v\in V(G)$ 
$$|\mean(S_{\FN'}(w):w\in N_v(G))-$$
$$\mean(1_\CX(w^\CX):w\in N_v(G))|<g_3(\epsilon')$$
Hence, by $\comb'^{(t+1)}$ being Lipschitz-Continuous, there exists 
$${g_4:\Real\rightarrow\Real, \lim_{x\rightarrow0}g_4(x)=0}$$
such that for every $(G,\Cb)\in\CGS^\bool_{p^{(1)}},v\in V(G)$
$$ |\comb'^{(t+1)}(S_{\FN'}(v),
    \mean(S_{\FN'}(w):w\in N_v(G)))-$$
    $$\comb'^{(t+1)}(1_{\CX}(v^\CX),\mean(1_{\CX}(w^\CX):w\in N_v(G)))|<g_4(\epsilon')$$
Assuming $\epsilon'$ small enough such that $\forall \epsilon''< \epsilon'\;\;g_4(\epsilon'')<\frac{\epsilon}{4}$ we have that for every $(G,\Cb)\in\CGS^\bool_{p^{(1)}},v\in V(G)$
\begin{equation}\label{eq:eq_6}
    \begin{split}
    &|\comb'^{(t+1)}(S_{\FN'}(v),\mean(S_{\FN'}(w):w\in N_v(G)))-
    \\&\comb'^{(t+1)}(1_{\CX}(v^\CX),\mean(1_{\CX}(w^\CX):w\in N_v(G)))|<\frac{\epsilon}{4}
    \end{split}
\end{equation}
Define $\FL'^{(t+1)}\coloneqq(\id_n,\mean,\comb'^{(t+1)})$ and redefine $\FN'$ to include $\FL'^{(t+1)}$, that is,
$$\FN'\coloneqq(\FL'^{(0)},\ldots,\FL'^{(t)},\FL'^{(t+1)})$$
Combining \Cref{eq:eq_3} with \Cref{eq:eq_5}, and \Cref{eq:eq_4} with \Cref{eq:eq_6}
we have that for every $(G,\Cb)\in\CGS^\bool_{p^{(1)}},v\in V(G)$ 
$$|S_\FN(v) - x_v|<\frac{\epsilon}{2}\;,\; |S_{\FN'}(v) - x_v|<\frac{\epsilon}{2}$$
Hence, define 
$$\CX^{(t+1)}\coloneqq \{\comb^{(t+1)}(y, \Msg^{(t+1)}_\CX(y)\cdot\vec{p}) : (y,\vec{p})\in \CX\times S_{n,\delta_1}\}$$
then $\CX^{(t+1)}$ satisfies the requirement.
\end{proof}

\end{document}
